\documentclass[10pt, conference, letterpaper]{IEEEtran}
\IEEEoverridecommandlockouts
\pdfoutput=1
\usepackage{cite}
\usepackage{breakurl}           
\usepackage{url}                
\usepackage{CJKutf8}
\usepackage{amsmath,amssymb,amsfonts}
\usepackage{algorithmic}
\usepackage{float}
\usepackage{graphicx}
\usepackage{textcomp}
\usepackage{xcolor}
\usepackage{tabularx}
\usepackage{xspace}
\usepackage{multirow}
\usepackage{subfigure}
\def\BibTeX{{\rm B\kern-.05em{\sc i\kern-.025em b}\kern-.08em
    T\kern-.1667em\lower.7ex\hbox{E}\kern-.125emX}}

\newcommand{\name}{AutoByte\xspace}

\begin{document}

    \title{Automatic Configuration for Optimal Communication Scheduling in DNN Training}
    \author{Yiqing Ma$^1$, Hao Wang$^1$, Yiming Zhang$^{2}$, Kai Chen$^1$\\$^1$iSING Lab, Hong Kong University of Science and Technology\\
$^2$National University of Defense Technology}

    \maketitle

\begin{abstract}
ByteScheduler partitions and rearranges tensor transmissions 
to improve the communication efficiency of distributed Deep Neural Network (DNN) training. The configuration of hyper-parameters (i.e., the partition size and the credit size)
is critical to the effectiveness of partitioning and rearrangement.
Currently, ByteScheduler adopts Bayesian Optimization (BO)
to find the optimal configuration for the hyper-parameters beforehand.
In practice, however, various runtime factors (e.g., worker node status and network conditions)
change over time, making the statically-determined one-shot configuration result suboptimal for real-world DNN training. 

To address this problem, we present a real-time configuration method (called \name) that automatically and timely searches the optimal hyper-parameters as the training systems dynamically change. \name extends the ByteScheduler framework with a meta-network,
which takes the system's runtime statistics as its input
and outputs predictions for speedups under specific configurations. Evaluation results on various DNN models show that 
\name can dynamically tune the hyper-parameters with low resource usage, 
and deliver up to 33.2\% higher performance 
than the best static configuration in ByteScheduler.

\end{abstract}

    \section{Introduction}
\label{sec:introduction}

Deep learning (DL) has been widely adopted for data mining and analytics. 
DL drives rapid development of emerging applications
such as face recognition~\cite{resnet2016cvpr} and natural language processing (NLP) \cite{chowdhury2003natural}. 
To support these applications, 
the datasets become larger and larger and the models become more and more complex~\cite{gpt3}, 
making DL training increasingly time-consuming. 
As a result, large-scale training is often conducted in a distributed manner. 
In distributed training, communication is usually the  bottleneck~\cite{pipedream}. 
Many studies have been done to improve the communication performance
for distributed training,
such as Parameter Server (PS)~\cite{ps-osdi}, All-reduce~\cite{ring}, and Gradient Compression~\cite{lin2017deep}. 

Most recently, communication scheduling~\cite{p3,tictac,bytescheduler}
is proposed to further improve the communication efficiency of distributed training.
The key idea of communication scheduling is 
to change the transmission order of different DNN layers, 
in order to better parallelize the communication and computing tasks. 
The state-of-the-art ByteScheduler~\cite{bytescheduler} 
provides a generic communication scheduling service for different DL frameworks 
(such as MXNet~\cite{mxnet}, PyTorch~\cite{pytorch} and Tensorflow~\cite{tensorflow}),
which adopts tensor partitioning and priority-based communication scheduling 
to realize pipelined communication and computation of distributed training.


Based on experimental analysis, 
we observe that 
the configuration of hyper-parameters (i.e., the partition size and the credit size) 
is critical to ByteScheduler
and inappropriate hyper-parameters may significantly degrade the effectiveness of partitioning and rearrangement.
Currently, 
ByteScheduler adopts Bayesian Optimization (BO)~\cite{brochuTutorialBayesianOptimization2010} to accelerate the searching process 
by finding a near optimal configuration of the hyper-parameters beforehand. 
BO assumes the optimal values of the hyper-parameters stay constant 
throughout the entire training procedure. 
In practice, however, various runtime factors, 
such as worker node status and network conditions, 
change over time, 
making the statically-determined one-shot configuration result suboptimal 
for dynamic training environments.


To address this problem, 
in this paper we present a real-time configuration method (called \name) 
that automatically and timely searches the optimal hyper-parameters 
as the training systems dynamically change. 
\name extends the ByteScheduler framework with a meta-network, 
which takes the system's runtime statistics as the model input 
and outputs predictions for speedups under specific configurations. 
The runtime statistics include static parameters 
(such as the numbers of workers and model types)
and dynamic parameters 
(such as layer-wise computation time and transmission rate 
which characterize the dynamic computation and communication conditions).
\name integrates offline training with online adaptation for its meta-network
to better fit the practical training scenarios. 
\name executes reconfigurations for the dynamically-generated optimal hyper-parameters 
based on an optimization trigger mechanism,
which can timely adjust the hyper-parameters 
and always achieve high training performance as runtime factors change.

Our evaluation on various DNN models shows that 
\name can dynamically tune the hyper-parameters to effectively improve the training performance. 
We show that 
the configuration automatically generated by \name 
performs very close to the real optimal configuration found by Grid Search
\cite{bytescheduler},
and that 
when multiple training jobs share the network bandwidth
(as in real cloud environments), \name can quickly find the optimal configuration under varying network bandwidth 
and improve the training performance by up to 19.2$\%$. 
We further evaluate \name's effectiveness for different bandwidth settings and communication architectures (PS and All-reduce),
and the results show that
\name can outperform the state-of-the-art communication scheduling algorithms 
by up to 33.2 $\%$
at low resource usage cost.

The rest of this paper is organized as follows. In $\S$\ref{sec:background}, we introduce the necessary background of communication scheduling in distributed training. We illustrate the downgrading caused by inappropriate partition size and credit size in ByteScheduler with both diagrams and experiments. Meanwhile, we show that the original hyper-parameters tuning technique Bayesian Optimization (BO) in ByteScheduler fails to adapt with the dynamic factors changing, e.g., available bandwidth and computation resource, which incurs suboptimal decisions. In $\S$\ref{sec:design}, we first explore the dynamic factors which impact the choosing of optimal credit size and partition size. Then we introduce the design of \name, at its core, \name uses a meta network optimizer to automatically exploit the relation of input metrics and training speed. To mitigate the extra overhead, \name leverages an offline training and online adapting strategy. In $\S$\ref{sec:evaluation}, we evaluate the speedup of \name compare to vanilla ML framework and ByteSheduler on three different models. We also show the rapid adaptation and low overhead of \name.

    \section{Background \& Motivation}

\begin{figure}[t]
\centering
\subfigure[Computation and communication in distributed deep learning architecture] {
	\includegraphics[width=0.45\textwidth]{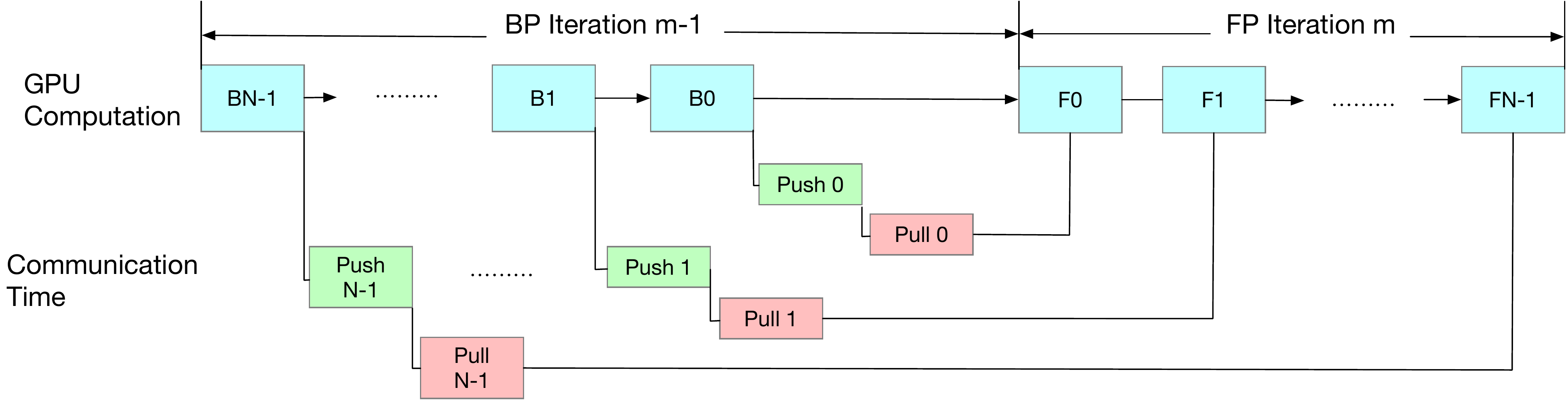}
	\label{figs:cc}
}
\subfigure[A contrived example showing performance gain with tensor partitioning]{
	\includegraphics[width=0.45\textwidth]{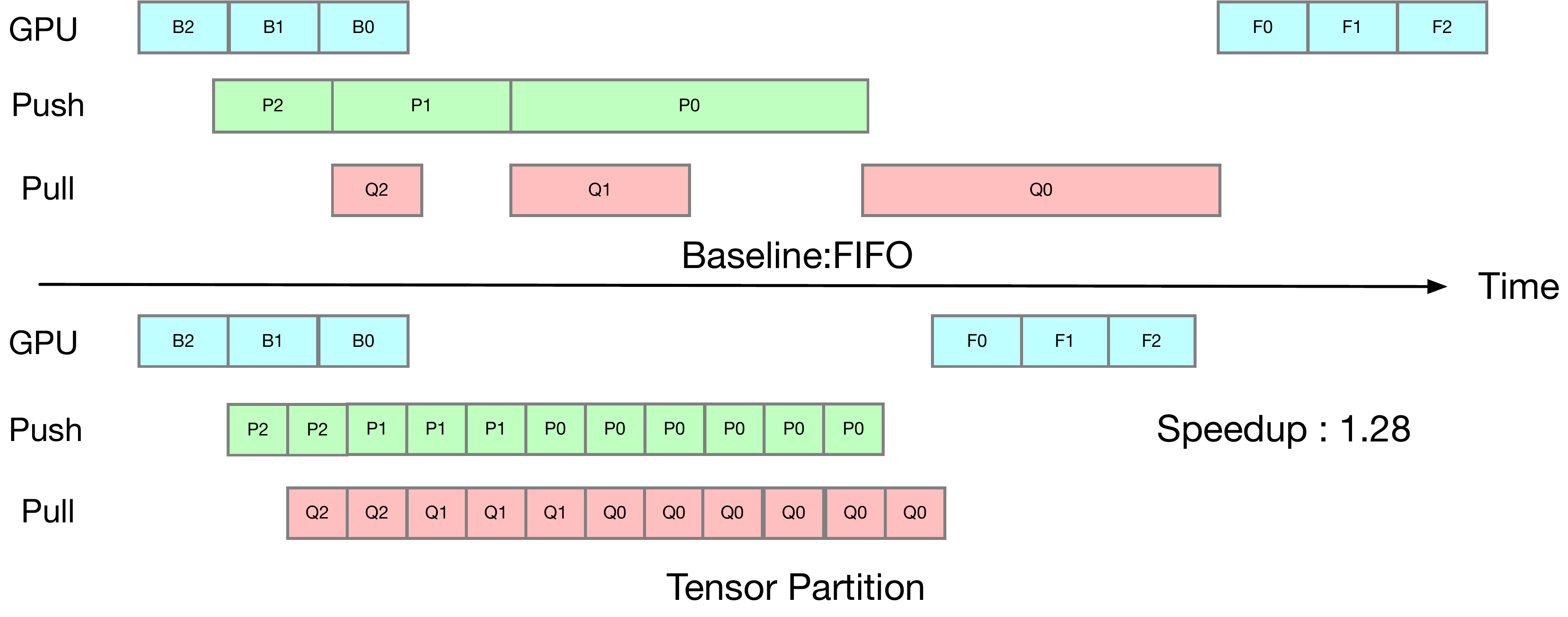}
	\label{figs:tp1}
}
\subfigure[tensor partitioning with inappropriate partition size will cause performance deficiency]{
	\includegraphics[width=0.45\textwidth]{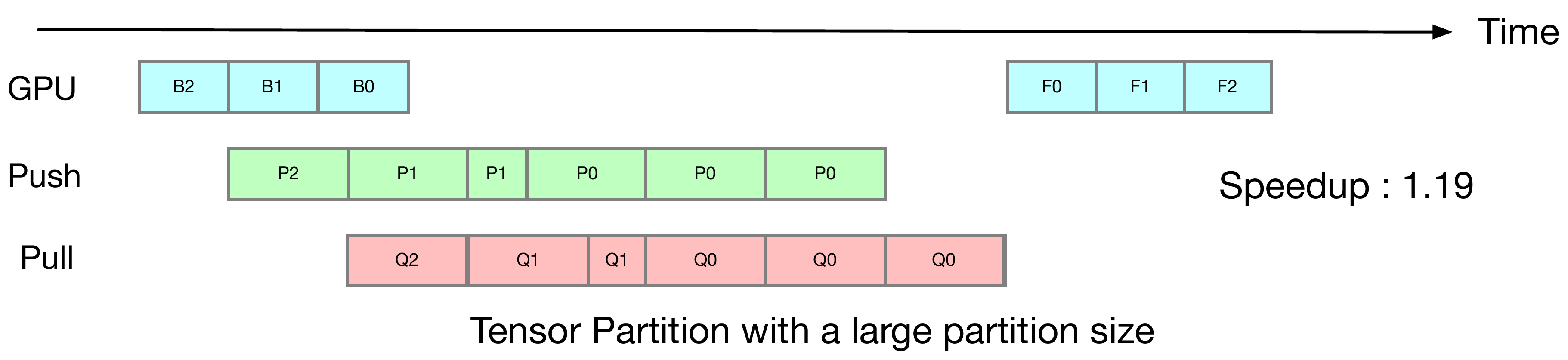}
	\label{figs:tp2}
}
\caption{Tensor partitioning. 
An appropriate partition size brings significant performance gains.}
\label{figs:tp}
\end{figure}


\begin{figure}[t]
\centering
\subfigure[Credit-based priority scheduling allow higher-priority tensors to jump ahead of the queue] {
	\includegraphics[width=0.45\textwidth]{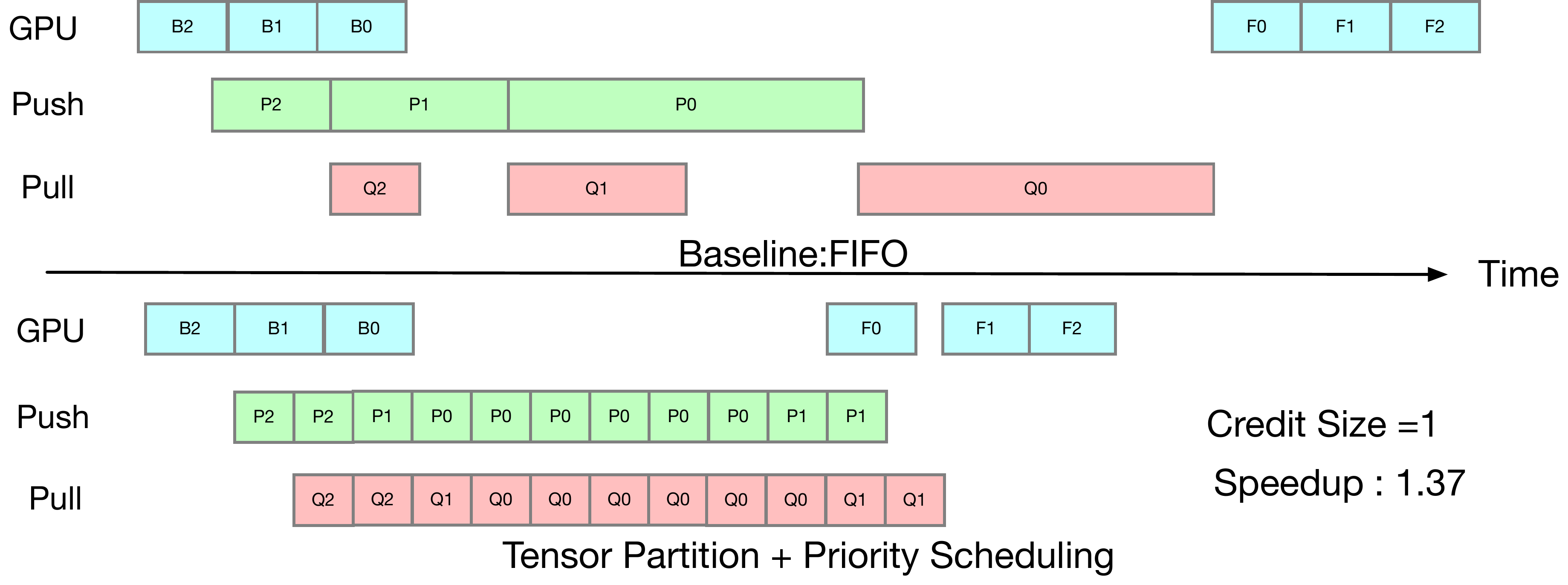}
	\label{figs:cs1}
}
\subfigure[An appropriate credit size can fully utilize the network bandwidth and deliver positive gains]{
	\includegraphics[width=0.45\textwidth]{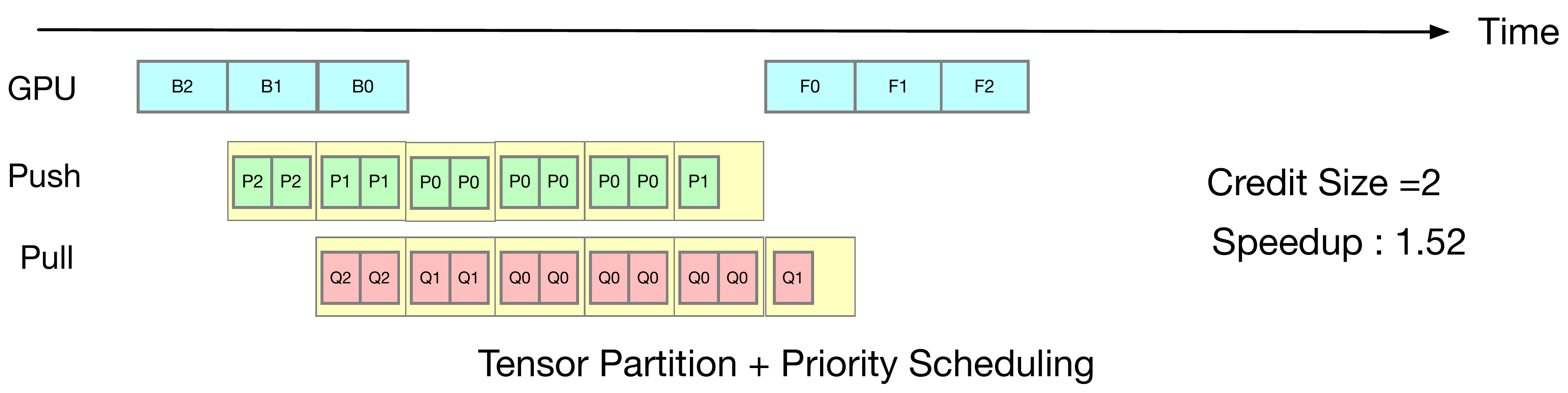}
	\label{figs:cs2}
}
\subfigure[Inappropriate credit size 
brings negative effect]{
	\includegraphics[width=0.45\textwidth]{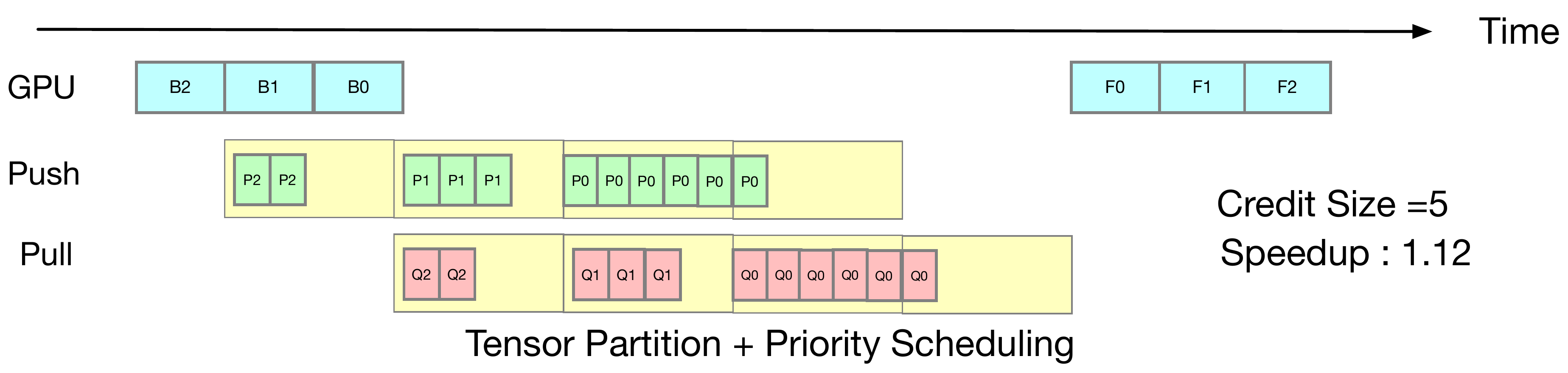}
	\label{figs:cs3}
}
\caption{Credit-based priority scheduling. 
An appropriate credit size also brings considerable performance gains.}
\label{figs:cs}
\end{figure}

\label{sec:background}
In this section, we introduce the background of distributed training. 
We describe the necessity of distributed training, the basic workflow, the common distributed training architectures
and classical communication scheduling strategies.

\subsection{Distributed Deep Learning}
\label{subsec:ddl}
For a deep learning task, a DNN model can be trained through an iterative way. The iterations are repeated over the large training dataset until the loss function is minimized and the model converges to an acceptable prediction accuracy~\cite{sgd}.

\par\noindent \textbf{Forward and backward propagation.} When dealing with the large training dataset, the dataset is firstly partitioned into many mini-batches~\cite{goyal2017accurate}. Then in each iteration, one mini-batch travels through the model layer-by-layer and generates a loss. This is the process of  \textit{forward propagation (FP)}. After that, the gradients are pushed through the last layer to the first layer. This is the process of \textit{backward propagation (BP)}. Lastly, the calculated gradients are used to update the model parameters by using Stochastic Gradient Descent (SGD)~\cite{sgd}. Then the model can start to deal with the next mini-batch similarly. Figure~\ref{figs:cc} shows the whole process.

\vspace{1mm}
\par\noindent \textbf{Data parallelism. } 
Due to the high degree of computational complexity in processing large amount of data and tuning billions of parameters, training such a DNN model is tremendously time consuming~\cite{goyal2017accurate}. These days distributed training is all the rage. Especially, data parallel distribution is a popular method for accelerating the training. The strategy of data parallelism is partitioning the dataset into multiple devices. Each device has the same model parameters but with different partitions of the dataset. In each iteration, the calculated gradients are aggregated to update the model parameters, and then broadcasted to all devices. 

\subsection{Communication Architecture}
\label{subsec:nca}
In the parameter update step, devices need to share their gradients, which involves network communication issues. Two kinds of communication architectures are widely used. They are Parameters Server~\cite{ps-osdi} and all-reduce~\cite{ring}.


\vspace{1mm}
\par\noindent \textbf{Parameter Server (PS).}
The PS architecture leverage parameter servers to collect and aggregate the gradients. Each worker computes the gradients locally and sends them to the servers through push step. Each server sums the gradients from different workers and update the parameters. Each worker then fetches the parameters from the servers through pull step. This architecture fully utilizes the computation power and is able to guarantee the synchronization of DNN training. A PS architecture also enables fault tolerance~\cite{ps-osdi} .


\vspace{1mm}
\par\noindent \textbf{All-reduce.}
Unlike PS, All-reduce does not require additional servers for the update of parameters. Instead, every worker collects the gradients from others' and update their parameters locally. One of the most popular all reduce methods is Ring All-reduce~\cite{ring}. A logical ring is formed among all workers. Every worker receives gradients from its left neighbor worker and sends the collected gradients to its right neighbor worker. Then they adds the gradients to its own copy until all gradients are aggregated. Then they broadcast the gradients along the ring. Ring all-reduce architecture can alleviate the bottleneck. It is also bandwidth optimal.

\subsection{Communication Scheduling}
\label{subsec:cs}
The layer-wise structure of deep learning training makes it convenient to parallel the communication tasks and computing tasks. The communication scheduling strategies mainly targets at minimizing the network communication time~\cite{lee2019automating,pipedream,poseidon,p3,tictac,bytescheduler,byteps,kang2020tensorexpress,wang2020domain,wang2020divide,xia2019rethinking}.
Poseidon~\cite{poseidon} supports overlapping communication process with backward propagation, reducing bursty network communication.
P3~\cite{p3} attempts to overlap communication with forward propagation by layer partitioning and scheduling on MXNet PS architecture. TicTac~\cite{tictac} proposes a similar idea that reduces the iteration time by calculating the order of transmission. It guarantees near-optimal overlap of communication and computation. It is only implemented on TensorFlow PS. ByteScheduler~\cite{bytescheduler} provides a generic communication scheduler for distributed DNN training. It supports TensorFlow, PyTorch, and MXNet on both Parameter Server and all-reduce architectures. PipeDream~\cite{pipedream} also exploits the scheduling on the model parallelism. it proposes to aggressively pipeline mini-batch processing to maximize the utilization of computing resources. However, its coarse-grained partitioning potentially misses optimization opportunities and may fail to scale extremely large layers. TensorExpress~\cite{kang2020tensorexpress} schedules tensor packets in-network using P4 to mitigate network contention and reduce training time. 
 
To sum up, there are two main techniques for communication scheduling, namely, tensor partition and priority-based scheduling~\cite{p3,bytescheduler}, which have remarkable effects and thus been widely used.
 
\vspace{1mm}
\par\noindent \textbf {Tensor partitioning.}
During the forward propagation, the computation of back layers must wait for the completion of front layers~\cite{poseidon}. However, during the communication, tensors of back layers are transmitted first since they finish earlier in the backward propagation~\cite{p3}. To tackle this mismatching, recent work propose tensor partitioning~\cite{bytescheduler}. At its core, a tensor will be divided into smaller chunks if its size is larger than a threshold, i.e, partition size. With the help of tensor partitioning, we are able to switch the transmission from back layers to front layers. The forward propagation can start immediately after receiving the front layers' tensors. Therefore, the overlap of the communication and computation time is increased and the training time is reduced. 

\vspace{1mm}
\par\noindent \textbf {Priority-based scheduling.}
As we mentioned above, forward propagation processes the front layers first, therefore, a straightforward idea is to prioritize the transmission of front layers. With the help of priority-based scheduling, we can set the priority to be the index of the tensor's layer. Thus these front-layer tensors can be transmitted first and the forward propagation can start earlier, which greatly mitigate the communication overhead. Priority-based scheduling is also proved to be the theoretical optimal~\cite{bytescheduler}.



\subsection{Hyper-Parameters for ByteScheduler} 
\label{subsec:pre}
There are two tunable parameters which are essential to the training performance: the partition size and the credit size~\cite{bytescheduler}.



\vspace{1mm}
\par\noindent \textbf{Partition size.}
 Figure~\ref{figs:tp} shows the effect of partition size.
 $S_{p}$ is the threshold of splitting the communication tensor. As shown in Figure~\ref{figs:tp1}, when the tensors are splitted into appropriate small pieces, the communication and computation are well overlapped, which results in a shorter training time,1.28$\times$ speedup. A larger partition size in Figure~\ref{figs:tp2} brings about performance deficiency(from 1.28$\times$ to 1.19$\times$). Intuitively, the ideal status is that the partition size is infinitesimal. Then the backward propagation and the former communication operation can start at the same time. This can lead to a shortest training time.

However, there is a gap between theoretical upper bound and the real world performance. The cost of tensor partition is not small enough to be ignored. This partition overhead will bring about additional time consumption, making infinitesimal partition size hard to satisfy in practice. There is usually an inherent partition size which can achieve optimal training speed~\cite{bytescheduler}.

\vspace{1mm}
\par\noindent \textbf{Credit size.}
Credit size $S_{c}$ is the size of a sliding window. It allows the tensor in the credit window size to be transmitted in parallel. The credit size is usually the multiple of the partition size. Intuitively, a larger credit size allows multiple tensors sending together. It allows the tensors filling the sending buffer in the network stack and fully utilize the network bandwidth~\cite{bytescheduler}.

As shown in Figure~\ref{figs:cs}. When the credit size equals to 1X in Figure~\ref{figs:cs1}, the priority scheduling policy is stop-and-wait since it allows one tensor in the sending buffer. Such a credit size is far from optimal in a 10Gbps network because of low bandwidth utilization. When the credit size equals to 2X as shown in Figure~\ref{figs:cs2}, the training time is reduced since a larger credit size leads to better bandwidth utilization. It increases the training speed from 1.37$\times$ to 1.52$\times$. However, a larger credit size will undermine the priority-scheduling. In Figure~\ref{figs:cs3}, when the credit size equals to 5X, the P0 tensors can not finish transmission before P1 tensors, thus F0 can not start beforehand. A larger credit size performs worse than a smaller credit size (the speedup drops to 1.12$\times$). Actually, credit-size is a critical parameter to balance the trade-off between the network utilization and priority scheduling.

\subsection{Motivating Examples}

\begin{figure}[t]
	\centering
	\subfigure[Model influence] {
		\includegraphics[width=0.42\textwidth]{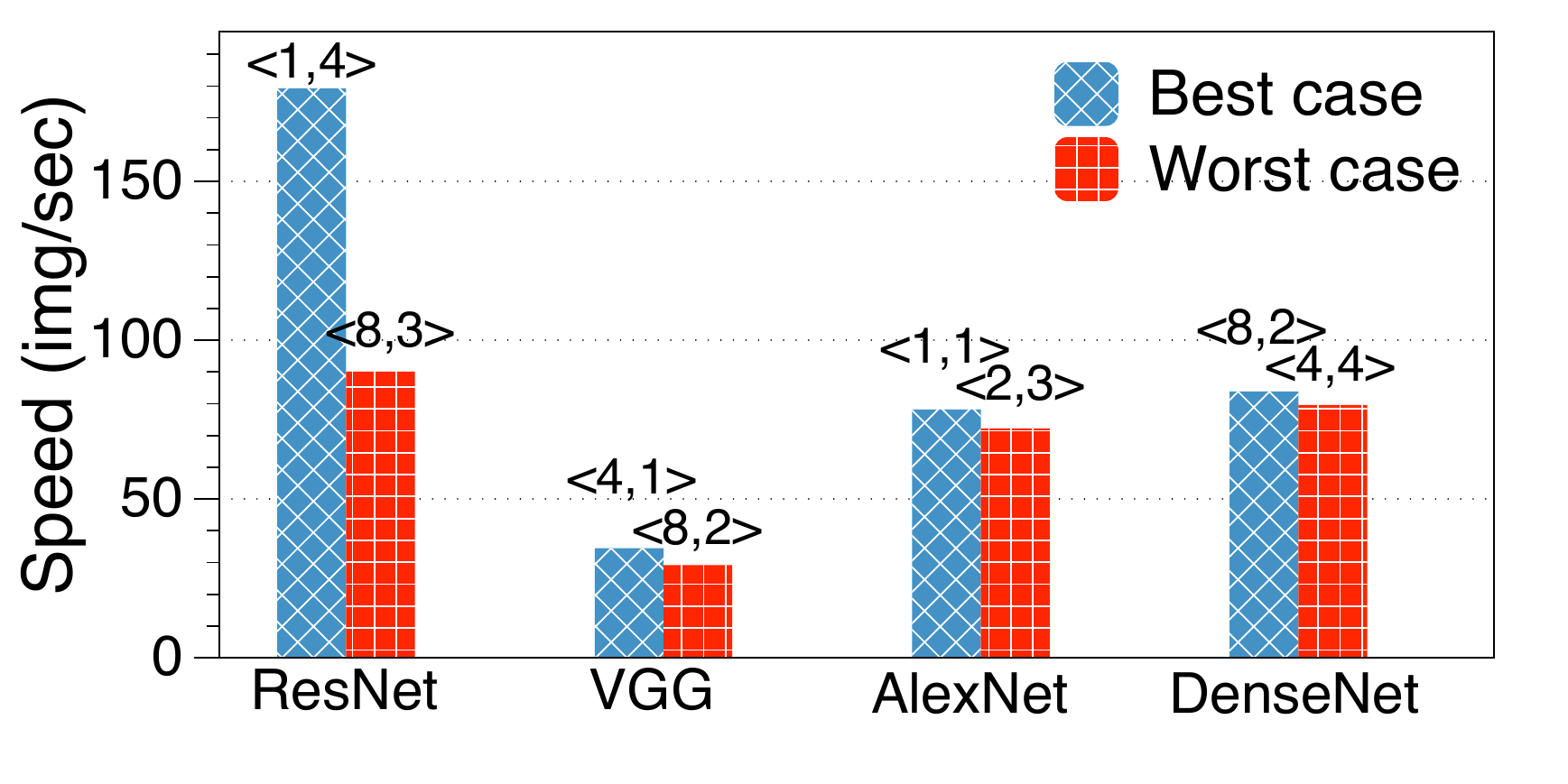}
		\label{figs:moti}
	}
	\subfigure[Bandwidth influence]{
		\includegraphics[width=0.42\textwidth]{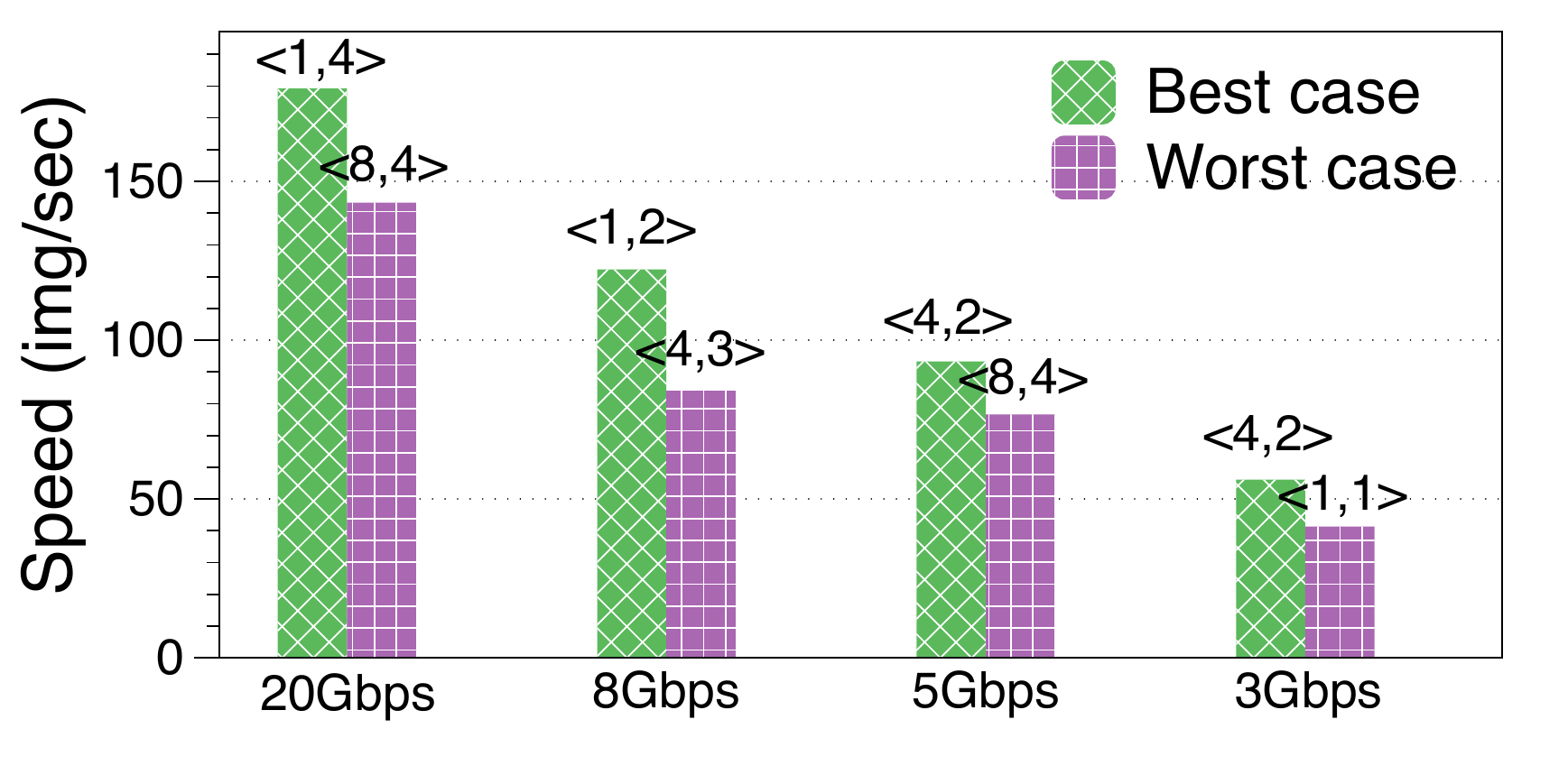}
		\label{figs:moti2}
	}
	\caption{Different models and available bandwidth have influence on optimal pair of $\langle S_p, S_c \rangle$.}
	\label{fig:motii}
	\end{figure}

These two important parameters, partition size and credit size, affect training performance. We use experiments to showcase this. We train four typical models, ResNet~\cite{resnet2016cvpr}, VGG~\cite{vgg}, AlexNet~\cite{alexnet} and DenseNet~\cite{densenet} with MXNet~\cite{mxnet} PS~\cite{ps-osdi} and PyTorch~\cite{pytorch} Allreduce~\cite{ring} (NCCL~\cite{nccl}). In the motivation experiments, we vary partition size from 1M to 8M and credit size from 1X to 4X. For the dynamic bandwidth experiment, we vary the bandwidth from 3Gbps to 20Gbps. The results are shown in Figure~\ref{fig:motii}.

The results show two points.
(1) Different ML jobs result in different optimal pair of $\langle S_p, S_c \rangle$. Different network environments do the same. As shown in Figure~\ref{figs:moti}, the optimal $\langle S_p, S_c \rangle$ is  $\langle 1, 4 \rangle$ for ResNet, 
    $\langle 4, 1 \rangle$ for VGG, $\langle 1, 1 \rangle$ for AlexNet and $\langle 8, 2 \rangle$ for DenseNet. Different ML models have respective proportion of communication to computation, thus their inherent parallel ratios are different. This reflects to different optimal configuration of $\langle S_p, S_c \rangle$.
    
    Besides, in Figure~\ref{figs:moti2} different bandwidth will have influence on the choice of optimal $\langle S_p, S_c \rangle$. The optimal  $\langle S_p, S_c \rangle$ is $\langle 1,4 \rangle$ for 20G bandwidth, $\langle 1, 2 \rangle$ for 8G bandwidth and  $\langle 4, 2 \rangle$ for 5G and 3G bandwidth. This is because in the real-world, scheduling and tensor partitioning have networking-related overhead. And credit-based priority scheduling is designed to fully utilize the network. Thus there exists an optimal pair of $\langle S_p, S_c \rangle$ in particular network setting.
 
 (2) Optimal pair of $\langle S_p, S_c \rangle$ can greatly improve the training speed. As shown in Figure~\ref{figs:moti}, the improvement of adjusting configuration ranges from $5.4\% - 99.0\%$ for different ML jobs and $21.8\% - 45.4\%$ for different network setting.

\begin{table}[]
\caption{Categorization of approaches on parameter tuning}
\centering
\resizebox{0.45\textwidth}{!}{
\begin{tabular}{|c |c |c |c |c |c |c |}
\hline & \multicolumn{2}{c|}{Dynamic} & \multicolumn{3}{c|}{Static} & \multicolumn{1}{c|}{Cost}\\ 
\hline Methods & Traffic & GPUs & Architecture  & Model & Network & Degree \\ 
\hline Default &  & &  &  && Low \\ 
\hline Grid Search  &  &  & $\checkmark$ &  $\checkmark$ &  $\checkmark$& High \\
\hline BO  &  &  &  $\checkmark$ &  $\checkmark$ &  $\checkmark$& Low \\
\hline \name &  $\checkmark$ &  $\checkmark$ &  $\checkmark$ & $\checkmark$ & $\checkmark$ & Medium\\
\hline
\end{tabular}}
\label{table:compare}
\end{table}

Actually, current communication scheduling methods~\cite{p3, bytescheduler} have used different methods to tune these two knobs $S_{p}$ and $S_{c}$. Table~\ref{table:compare} provides an overview and categorization of these relevant methods. 1)Default method. P3 ~\cite{p3} uses a default partition size of 160KB and credit size of 1X. This method has no additional overhead. But this partition size is far from optimal in 10Gbps network.
This stable method may lead to performance degradation in some particular network condition and ml models. 2) The Grid Search method. It is a simple way to search the best credit size and partition size. It enumerates all possible combinations of $\langle S_p, S_c \rangle$. But the cost of enumeration is too high. It is also ineffective where background traffic exists. 3) Bayesian Optimization (BO). ByteScheduler ~\cite{bytescheduler} adopts BO to tune $\langle S_p, S_c \rangle$ together. BO can reach optimal configuration with much less cost and give much more stable performance. However, BO only works before the beginning of training. Once the train starts, the $\langle S_p, S_c \rangle$ stay constant. It  can not deal with the dynamic condition like bandwidth competition and computing power change. 4) \name. Our proposed method should not only find the optimal pair for particular ML model and network, but also keep the optimal performance when sudden change happens.

	The optimal pair are likely to vary with many factors as listed in table~\ref{table:compare}. The physical network bandwidth, gradient synchronization method, DNN model types all have effects on these parameters. Sudden change of network bandwidth and computation power also have effects. Besides, the cost of reconfiguration should be within an acceptable range.	

\vspace{1mm}
\par\noindent \textbf{Design goal.}
The design of \name has three objectives.
\begin{itemize}
	\item \name should automatically choose the parameters to achieve optimal training performance in different run-time environments.
	\item \name should adjust the parameters when dealing with network or computation power change over the training courses to keep optimal performance.
	\item \name should have an acceptable search cost.
\end{itemize}
We will introduce how \name achieves the three objectives in $\S$\ref{sec:design}. 

    \section{Design}
\label{sec:design}
Through conducting the motivation experiments, we have demonstrated that how the DNN model and network bandwidth affect the optimal setting of the credit-size and partition size. Therefore, we extend an existing priority communication scheduling framework ByteScheduler to automatically configure the hyper parameters. We call the plugin as \name, which can adjust two core parameters:1) partition-size and 2) credit-size. We build a Meta Network Optimizer, which can predict the training speedup under different hyper parameter's configuration according to the runtime metrics, i.e., the computation power and background traffic bandwidth. Thus we can choose the optimal hyper-parameter setting and increase the training speedup.

 However, dynamic tuning the two parameters is non-trivial due to several reasons. Firstly, formally modeling and solving this problem is nearly impossible, since we can not explicitly express the training performance as the function of two parameters. Secondly, the distributed training system is very complicated. This is reflected in many places. For example, different DNN models vary in number of layers, each layer's computation time and the size of parameter.The tuning experience in one DNN model can not be directly applied to another. Next the computation power and communication bandwidth settings are vary in different runtime environment. Furthermore, it's quite complex to analyze particular cases such as computation straggling and communication straggling. 

To solve this problem, the first thought that comes to mind is to consistently search for the best values and use newly profiled results. Grid search and Bayesian Optimization(BO)\cite{brochuTutorialBayesianOptimization2010} are both able to search for the optimal partition size and credit size. But these method may incur significantly search costs. The cost may even outweigh the improvement of training performance. 

	
Therefore, we propose to design a meta-network. This network makes a prediction of training speed under different pairs of parameter settings. Thus we can select the optimal pair with the maximum training speedup at the cost of one inference.

\subsection{Meta Network Optimizer}\label{subsec:optim}
\begin{table}[]
\caption{Important Runtime Metrics (Input Features)}
\label{table:symbol}
\resizebox{0.48\textwidth}{!}{%
\begin{tabular}{|l|l|l|}
\hline
Symbol & Shape & Specification                               \\ \hline
$S_c$  & $1$   & Credit size                                 \\
$S_p$  & $1$   & Partition size                              \\
$n$    & $1$   & Total number of workers                     \\
$l$    & $1$   & Total number of backward propagation layers \\
$m$    & $m*1$ & Model type embedding vector
	\\
$arc$  & $a*1$ & Architecture type embedding vector 
    \\
$B_d$  & $n*1$ & Network download speed for all workers      \\
$B_u$  & $n*1$ & Network upload speed for all workers        \\
T      & $l*n$ & Layer-wise computation time for all workers \\
V      & $n*1$ & Training speed (image/s) for all workers    \\ \hline
\end{tabular}%
}
\end{table}

\par\noindent \textbf{Input runtime metrics.}
Our target is to learn the mapping from the current environment and the partition size, credit size to the model training speed. Besides this two key hyper-parameters, the credit size $S_c$ and the partition size $S_p$, \name need to collect runtime metrics and use them as inputs to the meta network optimizer. We provide a detailed specification of the input and output features in Table ~\ref{table:symbol}, .i.e., the total number of workers $n$, the total number of backward propagation layers $l$, the model type of current job $m$, the architecture type of current work $arc$.

	Since the network condition and computation power are actually ever-changing during the training process, we should choose appropriate metrics to represent the dynamic runtime environment. We measure local computation time and communication time at the end of every mini-batch. 
	Some background traffic may preempt the bandwidth. Thus we need an index to show the real-time bandwidth. So we choose the network upload speed for all workers $B_{u}$ and network download speed  $B_{d}$ to show the real-time network transmission speed. Since the bandwidth metrics may fluctuate, we apply a n-dimension vector. On the other hand, the computation power also needs to be monitored to prevent sudden machine breakdown. Here we use a group of parameters to represent the real-time computation power, namely the layer-wise backward computation time of all workers $T = [T_1, T_2, .... T_n]$. With these runtime metrics, the dynamic changes of allocated computation power from GPU clusters and the network bandwidth can be detected. 
	
We denote the $V$ as the training speed, the objective of the meta network is to learn the mapping
\begin{align}
f : (T, B, S_c, S_p)\rightarrow V,
\end{align}
under the different settings of $S_p$ and $S_c$.  Then we can choose the best pair of $\langle S_p, S_c \rangle$ to optimize the training speed.

\begin{figure*}[ht]
\centering
\subfigure[Meta Network Optimizer]{
	\includegraphics[width=0.6\textwidth]{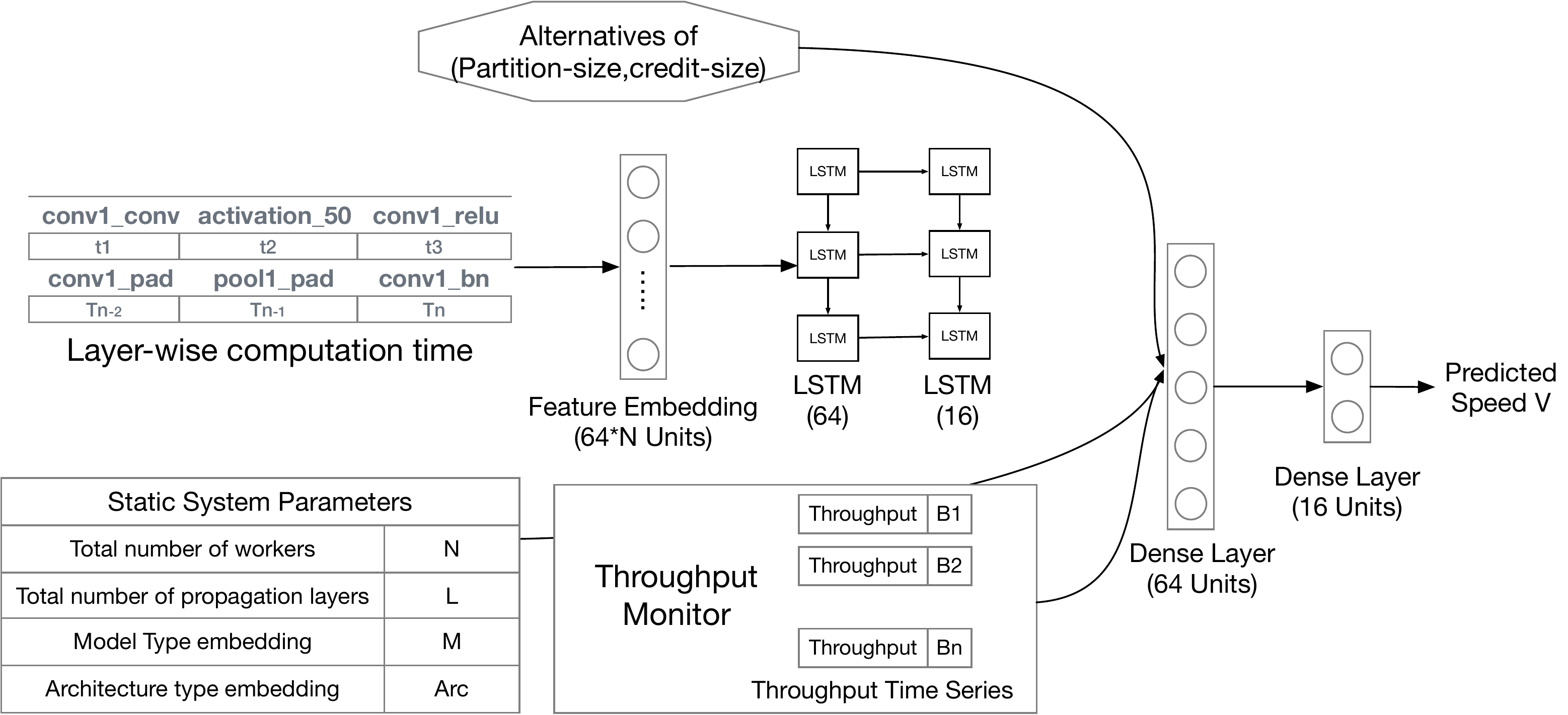}
	\label{figs:meta_network}
}
\subfigure[\name Workflow]{
	\includegraphics[width=0.34\textwidth]{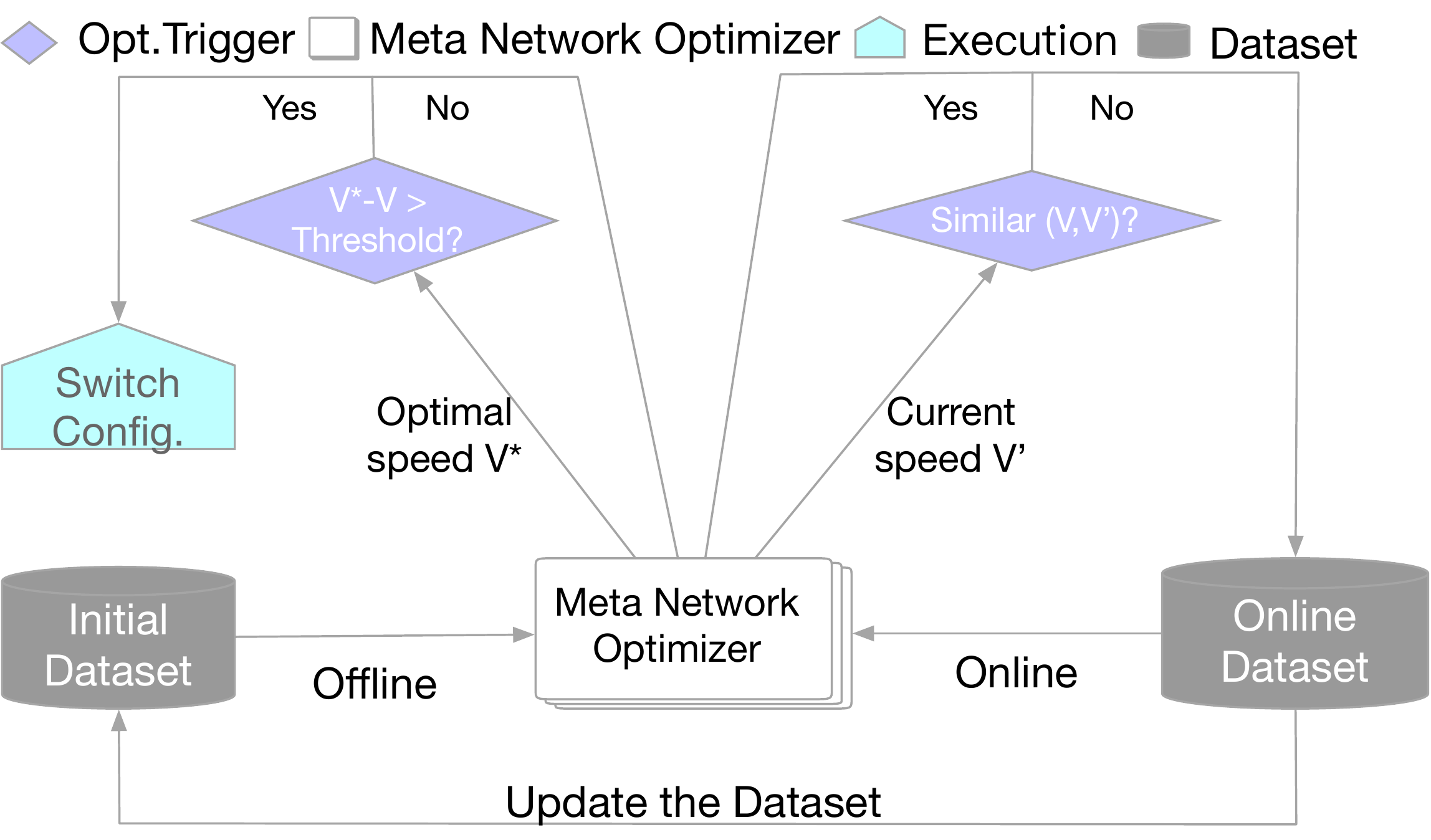}
	\label{figs:autos}
	}
\caption{The left figure illustrates the structure of Meta Network Optimizer. The right figure shows the workflow, and how Meta Network Optimizer Optimization Trigger and Execution work together.}
\end{figure*}



\vspace{1mm}
\par\noindent \textbf{Meta network architecture.}
The proposed meta network architecture is shown as Figure~\ref{figs:meta_network}. There are totally four  components. The first one is dynamic computation power monitor. As analysis before, we use layer-wise computation time to monitor the changing of computation power. It should be noticed that as different training tasks usually result in different number of layers. Therefore, we firstly embed layer-wise computation time $T$ into a fixed dimension feature space to make the meta-network robust to various of models. Afterwards, we apply two-layer LSTM to extract the sequential features in $T$. The second one is dynamic background traffic (network) monitor. We use real-time measured bandwidth series as $B_d$ $B_\mu$ to display network change. The third one is a static fixed-length parameter group, i.e. total number of workers. The fourth one is the alternative of $\langle S_p, S_c \rangle$. We concatenate the features extracted from the aforementioned four components. After applying two dense layers on the concatenated feature, the training speed $V$ is expected to be predicted. 

	As the objective of the meta-network is to predict precisely the speed up vector $V$. we define the loss function of the meta-network as a typical L2-loss function, i.e., 
	\begin{align}
      L(V, \overline{V}) = \|V - \overline{V}  \|_{L2}. 
	\end{align}
	Here, the $\overline{V}$ denote the observed average training speed. We set 10 iterations as a group. We calculate the average training speed for this group to eliminate the error and jitters.

\subsection{Training of Meta Network}
\label{subsec:train}

\par\noindent \textbf{Dataset collection.}
To train this meta network, the first step is to collect the training data. We use the testbed to run ByteScheduler to collect runtime data. We have implemented the meta-network using Keras. To evaluate the meta-network, we utilized ByteScheduler to train three typical models, ResNet50, VGG16 and AlexNet. For the partition size experiment, we vary the partition size from 4KB to 1024MB and credit size from 1X to 16X. To simulate the dynamic system conditions, we dynamically adjusted the network bandwidth in a wide range. We vary the bandwidth from 0.5, 1, 5, 10, 25Gbps at 20 iterations interval to simulate the coexisting with background traffic. we respectively collect each workers' runtime statistics and record the average training speed after 10 iterations. A total of 360000 iterations of meta-data samples have been collected. We can use the training data to develop our meta model \name. 

\vspace{1mm}
\par\noindent \textbf{Offline training online adapting.}
One critical issue is that DNN is a data-driven approach which can not handle the out of distribution problem. As the distributed deep-learning system environment may vary in both hardware(GPU and network) and software (model and DL architecture) settings, it is not practical to acquire a perfect distributed training dataset. 

One possible solution is to perform online training. we can  online train and update the meta-network on the target distributed deep learning task. However, this also introduces system overhead as we are required to try exhaustive parameter settings recurrently to meet with various of system conditions. This method violates our original intention, i.e.,maximizing the training speed. 

To overcome the issue, we proposed an offline training,  online adapting approach, whose key idea is use transfer learning to quickly adapt the meta-network to the current environment with low system overhead. This method has the following three advantages. 1)It does not bring about much additional system overhead. 2) Transfer learning helps the meta network optimizer quickly adapt to the current condition. 3) This method provides higher accuracy of speed prediction than offline trained version.

\subsection{\name workflow}
\label{subsec:workflow}
The workflow of the \name's online framework is shown in Figure~\ref{figs:autos}. \name contains three important components, Meta Network Optimizer, Optimization Trigger and Execution. The Meta Network Optimizer estimates the optimal configuration for the running ML job using runtime metrics. Following the decision of Optimization Trigger, the Execution applies the necessary changes dynamically to partition size and credit size without stopping the running job. 

\vspace{1mm}
\par\noindent \textbf{Meta network optimizer.}
When \name is deployed, for the Meta Network Optimizer, it collects runtime statistics and turn those metrics to vector inputs. Then the meta network will predict the speedup performance in different system configurations based on runtime metrics. After finding the configuration that is expected to be optimal, the Optimizer will generate a better configuration. Through reconfiguration, \name changes the system configuration to the one with better performance. But we need to make decisions such as when to calculate a better configuration, whether to execute the reconfiguration or not. To say that, we need an Optimization Trigger.

\vspace{1mm}
\par\noindent \textbf{Optimization trigger.}
In order to avoid the system from continuously reconfiguring back and forth around the estimated optimum. Optimizer Trigger predicts the performance benefit of a new configuration and skips that attempt if the gain is less than a certain threshold. A threshold number from our experience 5$\%$ is good enough to prevent the system from oscillating, while allowing the system to undergo moderately-sized optimizations.

The Optimization Trigger will also monitor the difference between the predict $V$ and the true speedup. When the difference is larger than a certain threshold $10\%$. It will trigger the online adapting scheme. The Meta Network Optimizer opportunistically tries to adapt the model to the current environment. It will use transfer learning to update the offline trained model and also update the dataset. Thus the optimizer can adjust to find an optimal configuration including the new online dataset.

\vspace{1mm}
\par\noindent \textbf{Execution.}
The execution of the optimization is relatively simple. The Meta Network Optimizer executes a plan by simply invoking the Execution API that reconfigures system transparently without stop training. Once the decision of a reconfiguration is made, the execution will switch the configuration of $\langle S_p, S_c \rangle$. Each time when the configuration is changed, we checkpoint and restart the training. The $S_p$ and $S_c$ can be modified by substituting the execution code.

    \section{Evaluation}\label{sec:evaluation}

\subsection{Methodology}\label{subsec:meth}

\par \noindent \textbf{Testbed setup}: Our testbed has 8 nodes, each with 20 CPU cores, 64GB memory, 1 Tesla V100, and a Mellanox CX-5 single-port NICs. We use a leaf-spine topology with 2 racks and 4 nodes in each rack. We use Mellanox SN2100 switch with the Onyx 3.7.1134 operating system. The link bandwidth is 100Gbps and the oversubscription ratio is 1. Our operation system is Ubuntu 18.04 with Linux kernel version 4.15.0-135-generic. The Mellanox driver version is 4.6-1.0.1.1. 

\vspace{1mm}
\par \noindent \textbf{Benchmarks}: We choose 3 typical models, ResNet50~\cite{resnet2016cvpr}, VGG16~\cite{vgg} and Alexnet~\cite{vgg}. We run our \name in MXNet~\cite{mxnet} with the PS~\cite{ps-osdi} and PyTorch~\cite{pytorch} with NCCL~\cite{nccl} (Ring all-reduce~\cite{ring}). For the PS, workers and servers are on the same machines, and their numbers are equal. We measure the performance under different NIC bandwidth. We chose TCP as our transport protocol. Since TCP cannot saturate the link bandwidth, we only test on 0.5, 1, 5, 10, 25 Gbps. We leave the RDMA test as the future work.

\vspace{1mm}
\par \noindent \textbf{Compared Schemes}: 
We compare \name with the following two different communication scheduling methods.
\begin{itemize}
	\item \textbf{Baseline}: We use the vanilla PS for MXNet and NCCL for PyTorch. For illustration, the backward propagation and forward propagation are executed sequentially without tensor partitioning and priority-scheduling. 
	\item \textbf{ByteScheduler}: ByteScheduler~\cite{bytescheduler}  is a generic communication scheduler, which combines priority scheduling and tensor partitioning. We use the open-source code of ByteSchedulerfor evaluation. We use two methods to set the partition size and credit size, namely, grid search and Bayes optimization~\cite{brochuTutorialBayesianOptimization2010}.
\end{itemize}

To clearly show how \name's Meta Network Optimizer works, especially how it optimizes the parameter configuration when network resource changes and how much overhead it incurrs. We first compare the Meta Network Optimizer with other methods, i.e., the Grid Search and Bayesian Optimization (BO). We use the code of BO method in ByteScheduler. We compare these three optimizers on the speed and the cost of reconfiguration. We measure how much iterations they need to find the optimal pair of $\langle S_p, S_c \rangle$,  and how much extra overhead they bring.

We use training speed (samples/sec) as the main metric. We measure the average training speed over the first 50 iterations after a warm-up of 10 iterations.

\subsection{Overview of the experiments}\label{subsec:summary}

We conduct the evaluation of \name and illustrate the result in the following four sections:
(1) In $\S$~\ref{subsec:validation}, we validate that \name can timely search for the optimal hyper-parameters during the training. We firstly introduce the method we use to find the optimal configuration. Then we show that \name can quickly find a nearly optimal hyper parameters.
(2) In $\S$~\ref{subsec:dynamic}, we evaluate \name when available bandwidth and computing resources are changing, which are common in a cloud environment. We emulate the resources changing by artificially adding a new training job. The experiment shows that \name outperformances ByteScheduler with BO by $18.2\% \sim 39.3\%$. 
(3) In $\S$~\ref{subsec:speed}, we show the speedup of \name in different environments. We compare \name with the baseline, ByteScheduler (with BO), on three models under different network bandwidth, different architectures (PS and Ring all-reduce). The result shows that with the help of Meta Network Optimizer, \name can improve the training speed with up to 94.4$\%$ and 33.2$\%$, compared with baseline and ByteScheduler. 
(4) In $\S$~\ref{subsec:deep}, we investigate the overhead introduced by \name. We also compare the reconfiguration speed of \name with other searching methods, namely, BO and grid search. 

\subsection{\name finds the Optimal Hyper-parameters}
\label{subsec:validation}

In this section, we evaluate \name to show that: 1) The output of \name can be very close to the optimal. 2) The hyper-parameters chosen by \name can fast converge to the optimal. 3) \name can bring considerable speedup to the training.

\begin{figure}[t]
	\centering
	\includegraphics[width=0.45\textwidth]{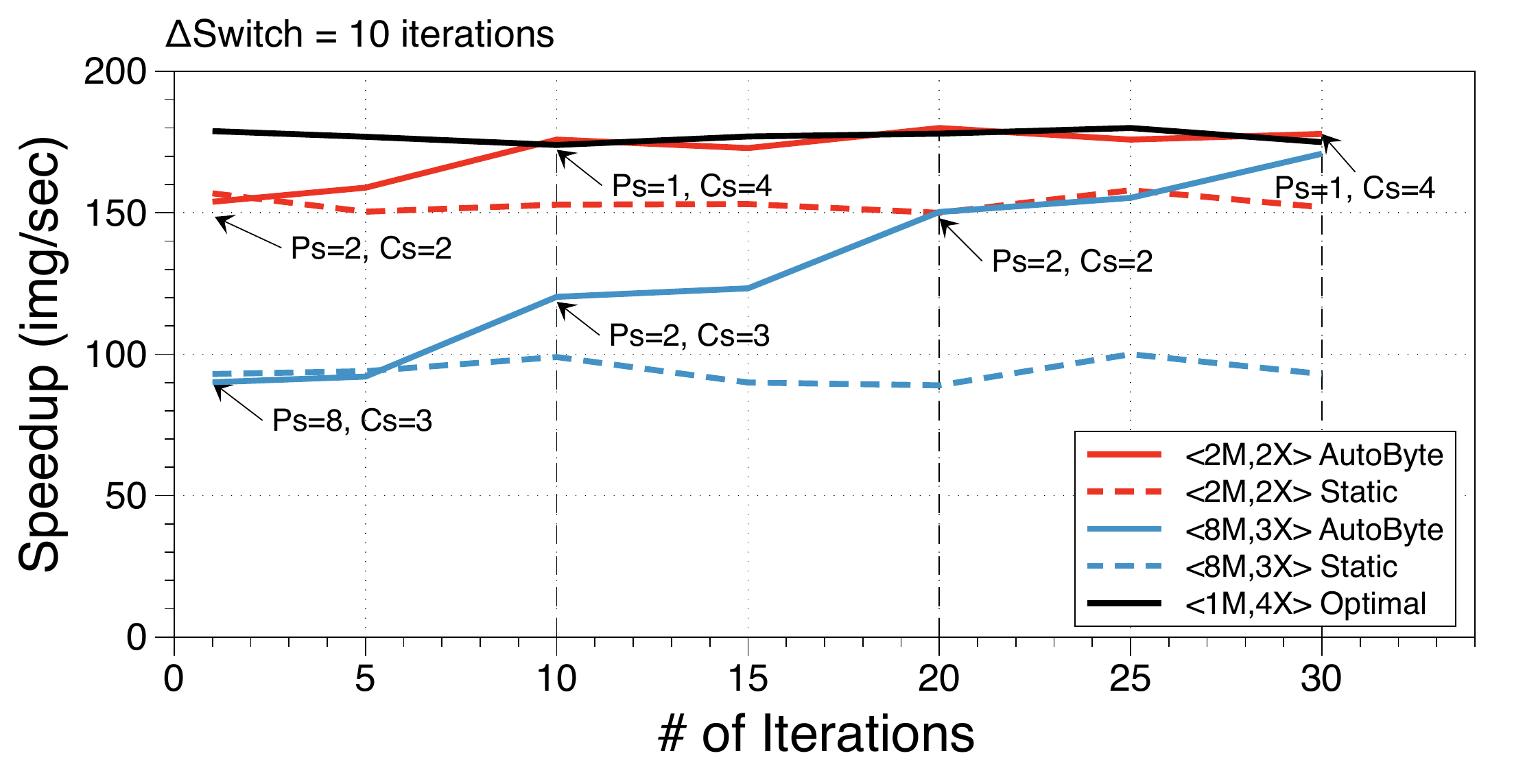}
	\caption{Training ResNet50 with 2 different configurations. The black line indicates the optimum, the red line and the blue line show two different configurations. The dotted lines show the performance without optimization. The vertical lines represent the reconfiguration. }
	\label{fig:switch}
\end{figure}

\begin{figure}[]
	\centering
	\includegraphics[width=0.45\textwidth]{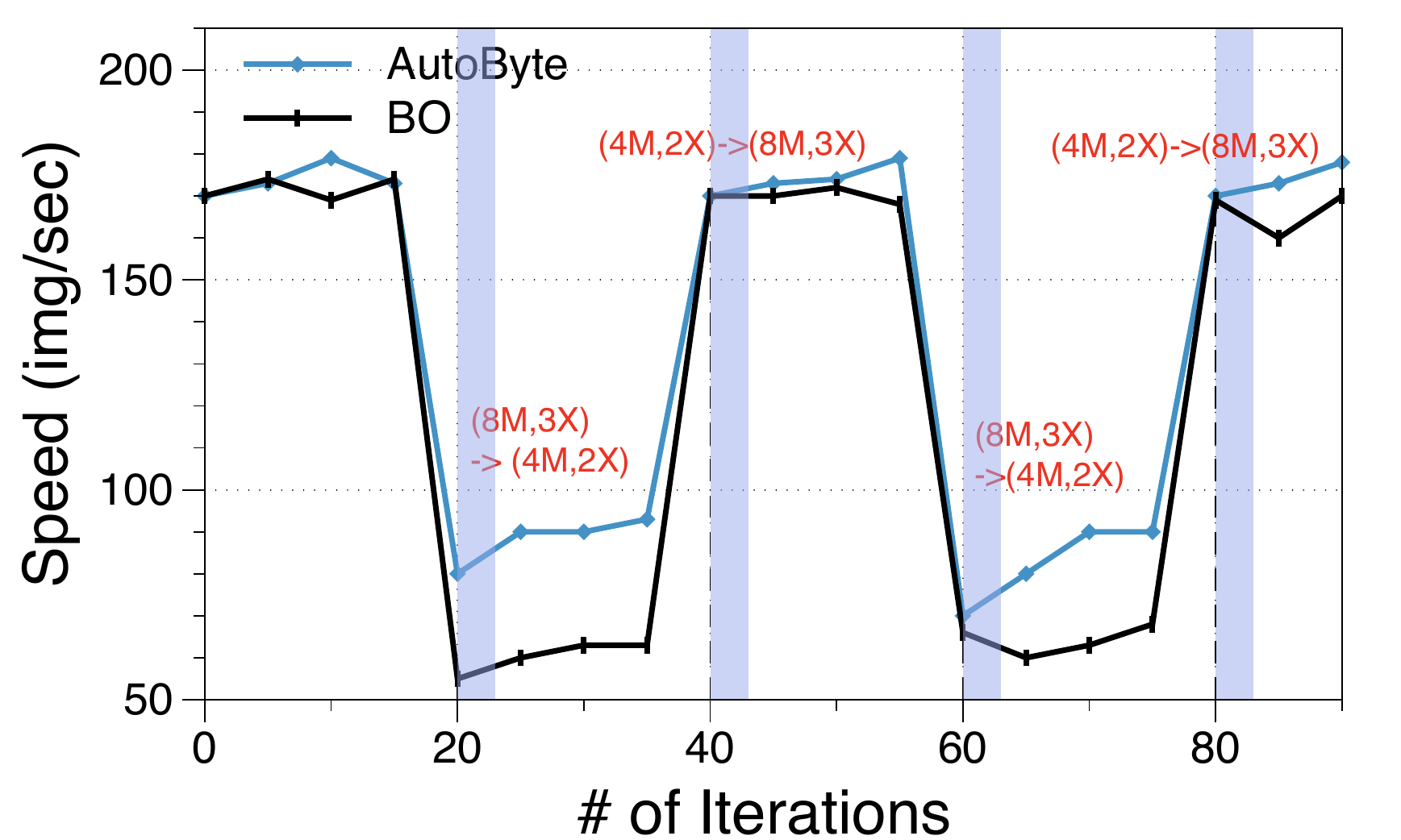}
	\caption{Training speed under dynamic bandwidth condition. Blue line shows \name's ability to adapt to dynamic network compared to BO drawn in black line, which assume the values stay constant throughout the training. The areas filled in sky-blue denote the reconfigurations. }
	\label{fig:dynamic}
\end{figure}

\begin{figure}[]
	\centering
	\includegraphics[width=0.45\textwidth]{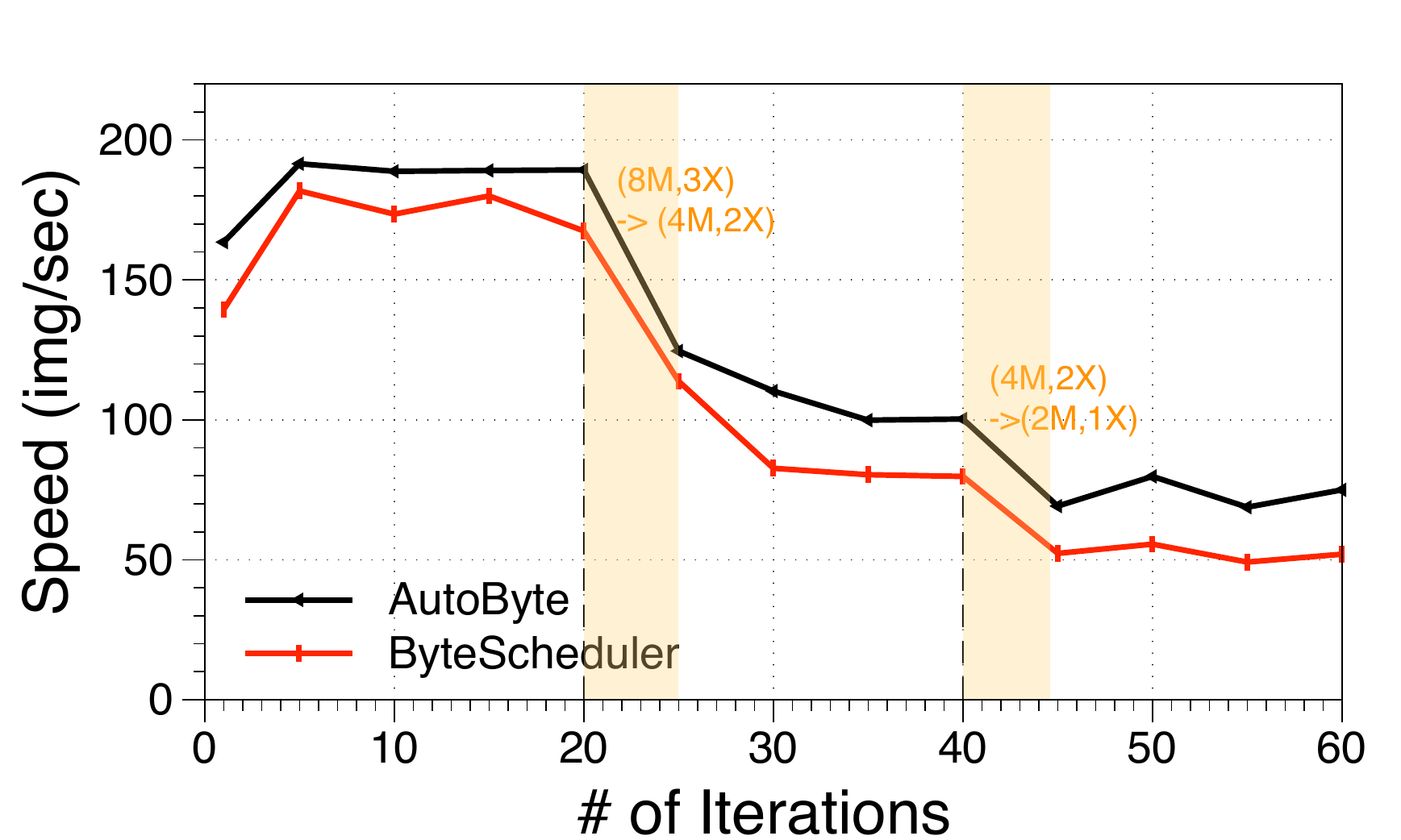}
	\caption{Training speed under dynamic changing of available resources. Black line and red line indicates the training of \name and ByteScheduler with BO, repectively. Vertical lines and areas filled in orange represent the event of reconfiguration.}
	\label{fig:dynamic2}
\end{figure}


\begin{figure*}[t]
\centering
\subfigure[ResNet50, PS, MXNet]{\includegraphics[width=0.31\textwidth]{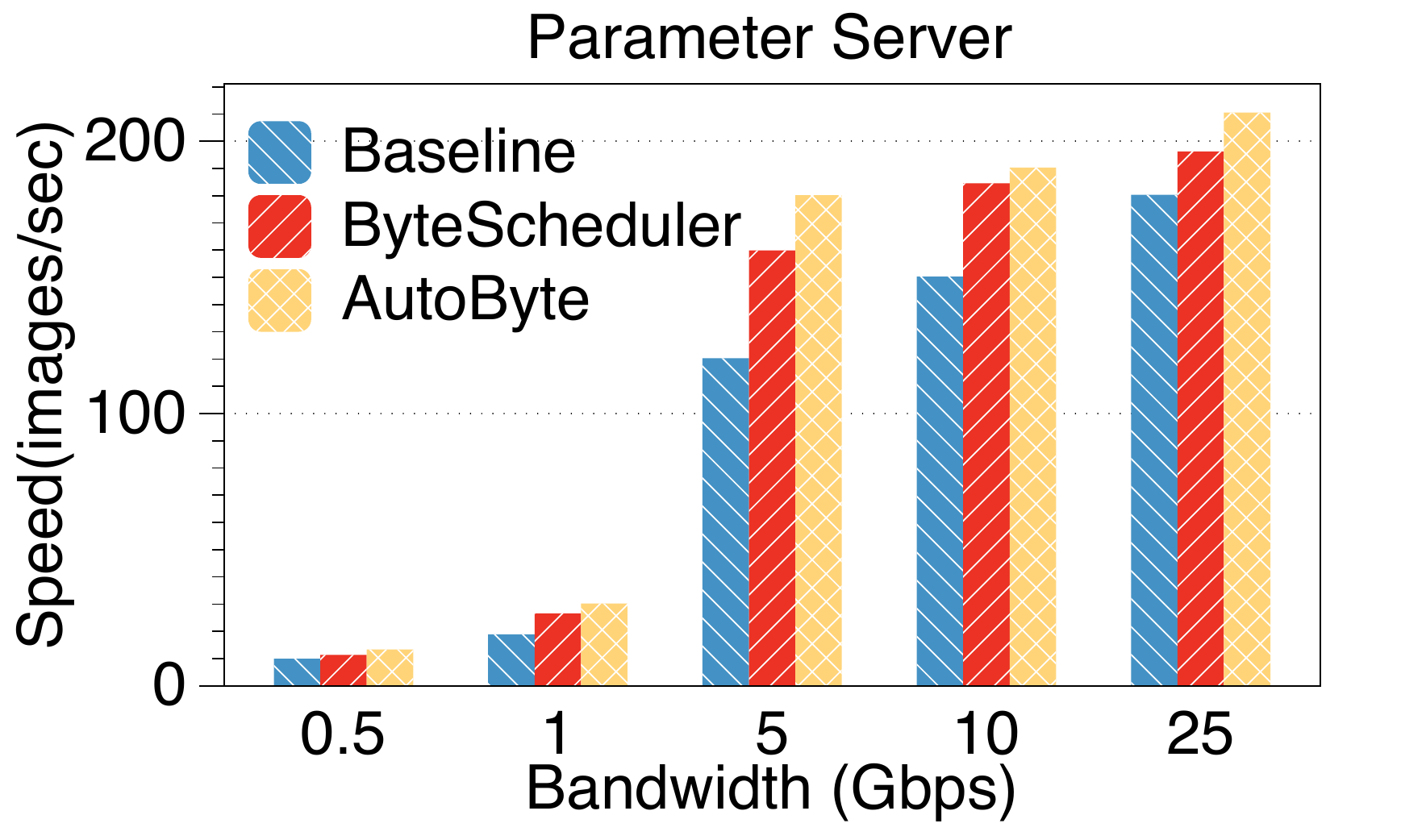}}
\subfigure[VGG16, PS, MXNet]{\includegraphics[width=0.31\textwidth]{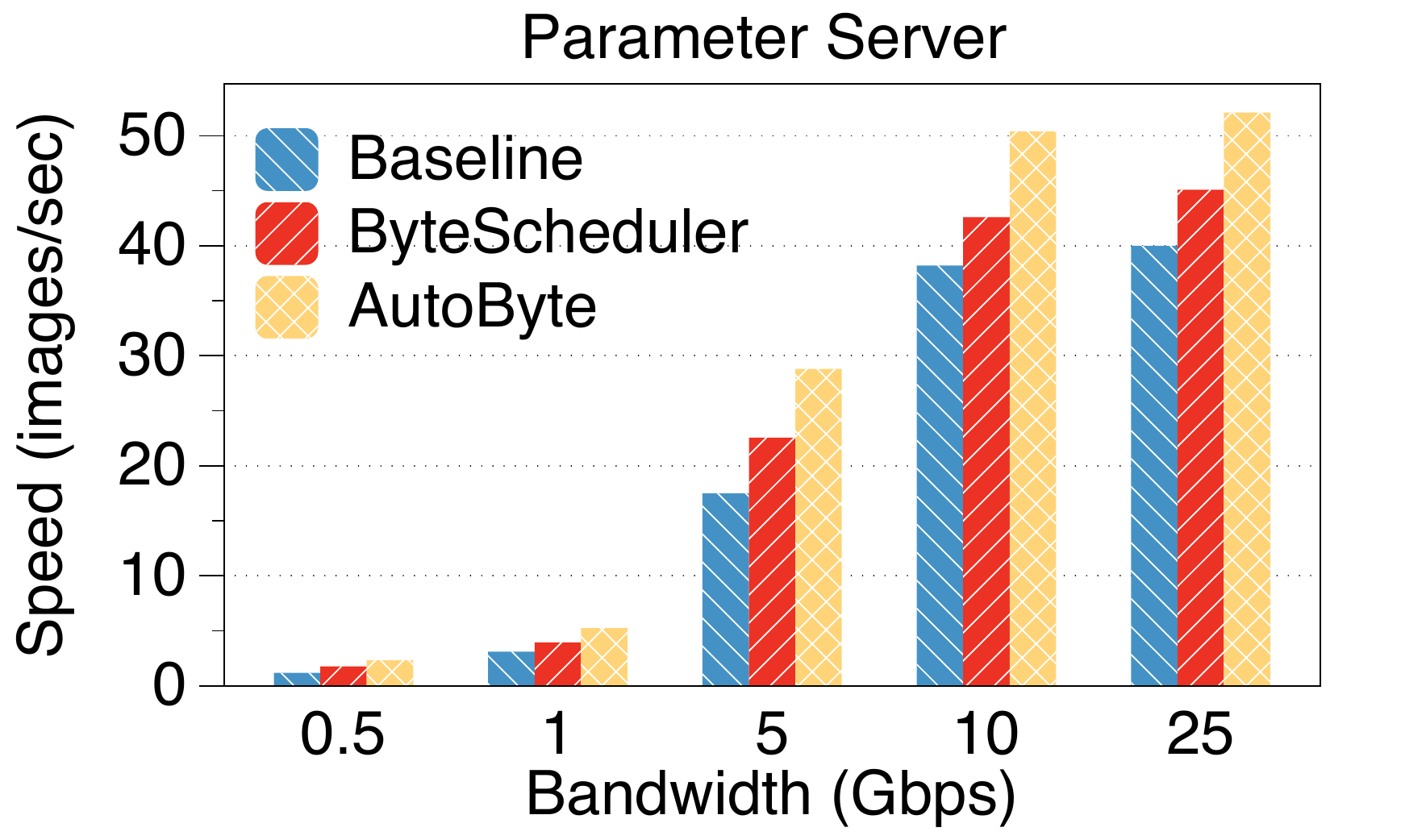}}
\subfigure[AlexNet, PS, MXNet]{\includegraphics[width=0.31\textwidth]{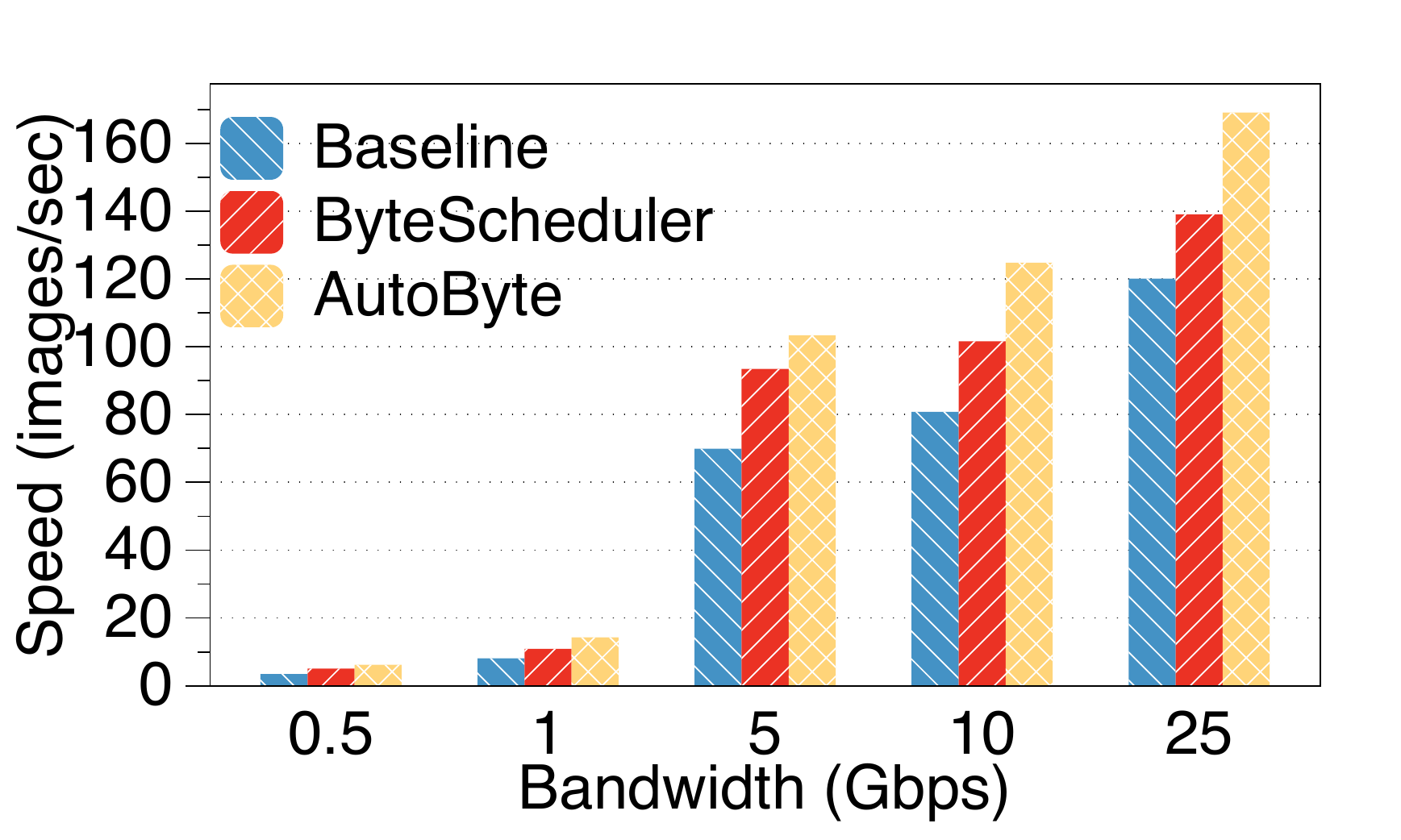}}
\caption{The training speed of ResNet, VGG, AlexNet models under different bandwidth conditions in Parameter Server.}
\label{fig:bw}
\end{figure*}

\begin{figure*}[t]
\centering
\subfigure[ResNet50, Ring, PyTorch]{\includegraphics[width=0.31\textwidth]{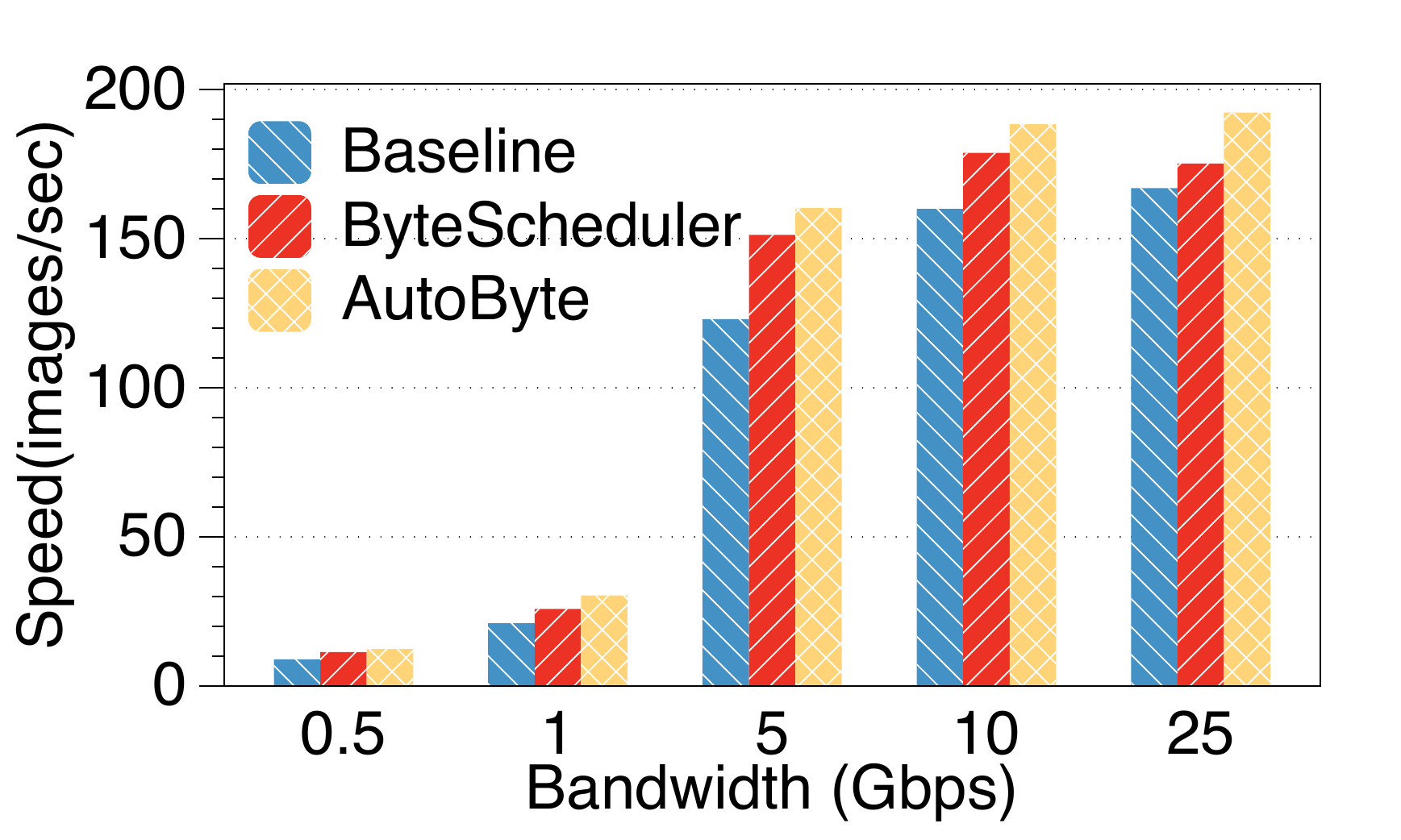}}
\subfigure[VGG16, Ring, PyTorch]{\includegraphics[width=0.31\textwidth]{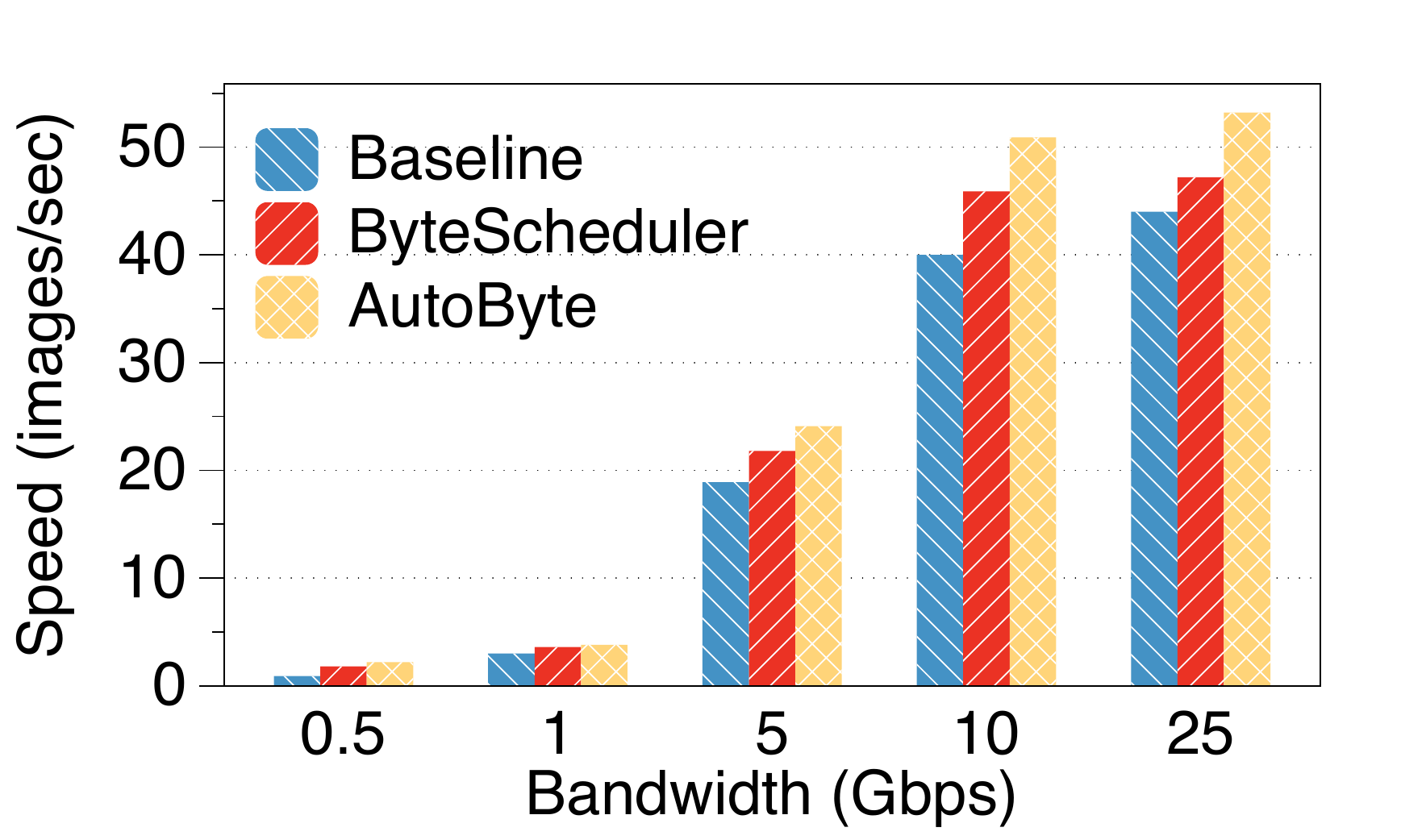}}
\subfigure[AlexNet, Ring, PyTorch]{\includegraphics[width=0.31\textwidth]{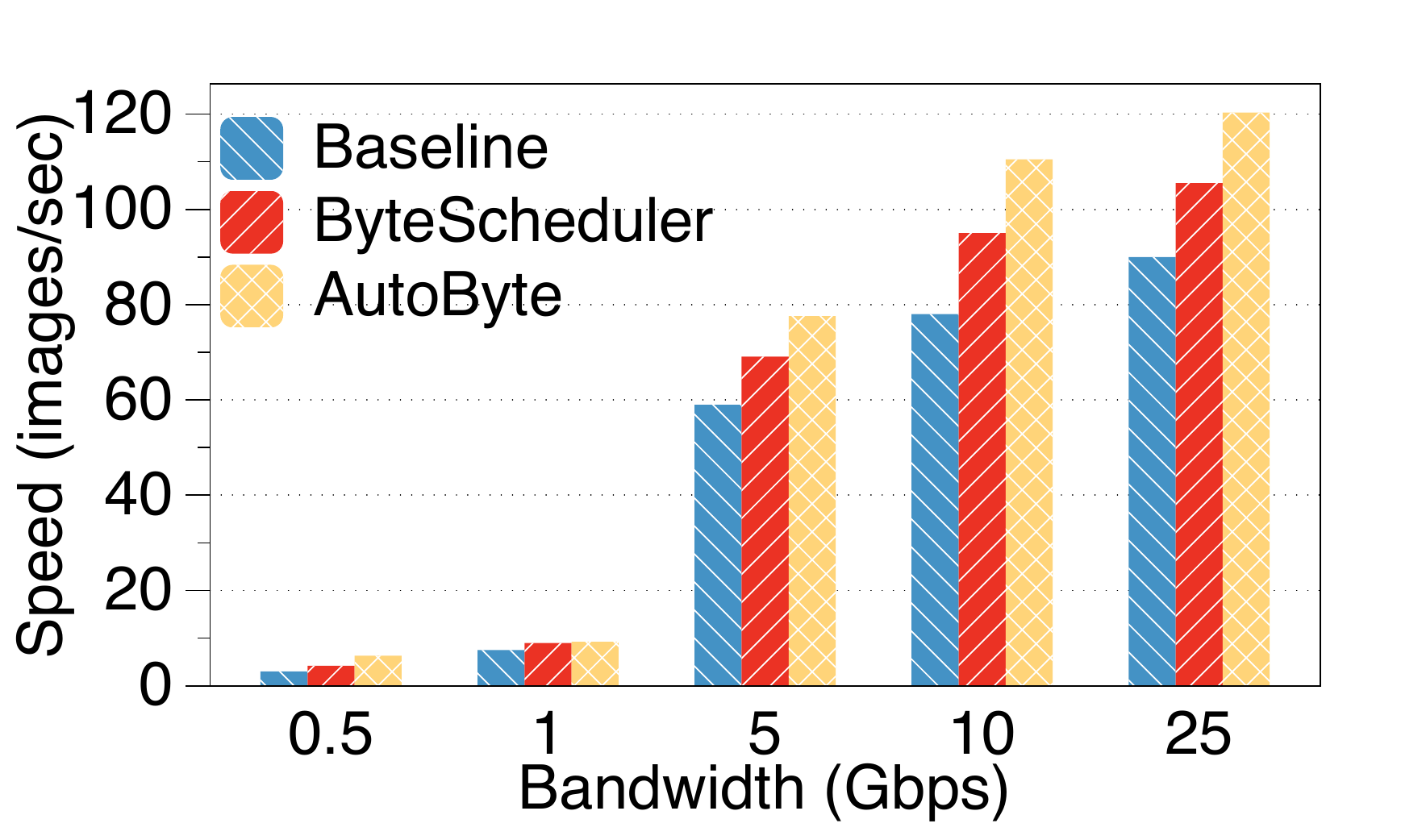}}
\caption{The training speed of ResNet, VGG, AlexNet models under different bandwidth conditions in All Reduce.}
\label{fig:bwring}
\end{figure*}

\vspace{1mm}
\par \noindent  \textbf {Optimal prediction.} We can obtain the optimal setting of $\langle S_p, S_c \rangle$ by grid search.  We first apply grid search to all possible configurations $\langle S_{p_{i}},S_{c_{i}}\rangle$, to find the ground truth. The step for partition size ($S_{p_{i}}$) is 2M, and for credit size ($S_{c_{i}}$) is 1X in grid search.  We denote the optimal pair as $\langle S_{p_{*}},S_{c_{*}}\rangle$. Figure~\ref{fig:switch} shows that the Meta Network Optimizer could predict the optimal configurations very close to the ground truth even with different initial parameters. 


\vspace{1mm}
\par \noindent \textbf {Fast convergence.} 
In our setting, \name conducts new prediction every 10 iterations.
Figure~\ref{fig:switch} shows that the output of \name can converge to the optimal after three times prediction, which is corresponding to 30 iterations, far less than the total iteration number in one epoch (often more than one hundred). Therefore, in practice, \name can quickly find the optimal configuration.

\vspace{1mm}
\par \noindent \textbf{Training improvement.} As shown in Figure~\ref{fig:switch}, for the red line, when change from  $\langle S_p\colon2, S_c\colon2 \rangle$ to  $\langle S_p\colon1, S_c\colon4 \rangle$, \name speedup the training by 12.6$\%$. For the blue line, when change from $\langle S_p\colon8, S_c\colon3 \rangle$ to  $\langle S_p\colon1, S_c\colon4 \rangle$, the speedup is 81$\%$.

\subsection{Optimization under Dynamic Resources}\label{subsec:dynamic}

The above experiments are under a static environment. Actually, \name can further speedup the training with a dynamic environment. \name can keep track of dynamic network and computation resources in the cluster and timely updates the parameter configuration. 

In the first experiment, we train a ResNet50 model. We simulate throughput variation through artificially changing the NIC bandwidth. We switch the bandwidth between 3Gbps and 10Gbps every 20 iterations. At the beginning, the bandwidth is 10Gbps.
 We compare the training speed of \name with ByteScheduler (BO). BO chooses the best credit size and partition size for the initial environment. Once chosen, it assumes the value stay constant throughout the training process. Figure~\ref{fig:dynamic} shows the average speed of training for every 5 iterations. At the $20^\mathrm{th}$ iteration, \name's configuration moves toward $\langle S_p\colon4, S_c\colon2\rangle$. This configuration better fits the low bandwidth setting, achieving higher speedup than ByteScheduler. At the $40^\mathrm{th}$ iteration, \name's configuration returns to the previous choice $\langle S_p\colon8, S_c\colon3\rangle$. Thus the \name can achieve 6.4$\%$-23$\%$ improvement in the training speedup.
 
In the second experiment, we simulate the changing of available bandwidth and computation resources by adding new training jobs. We add a new training job at $5^\mathrm{th}$ and $25^\mathrm{th}$ iterations. Since the job initialization takes 15 iteration's time, in the first $20^\mathrm{th}$ iterations, the measured job occupy the whole bandwidth (20Gbps) and the whole GPU. In the next 20 iterations, two jobs share the resources. In the last 20 iterations, three jobs share resources. 
As shown in Figure~\ref{fig:dynamic2}, we measure the average training speed of every 5 iterations. At the $20^\mathrm{th}$ iteration, \name's configuration moves toward $\langle S_p\colon4, S_c\colon2\rangle$. This configuration better fit the low bandwidth setting. Thus achieve higher speedup than ByteScheduler with BO. At the $40^\mathrm{th}$ iteration, \name's configuration turns to the $\langle S_p\colon2, S_c\colon1\rangle$. and \name achieve $8.8\% \sim 19.2\%$ higher performance. 

\subsection{Optimization under Various Environments}\label{subsec:speed}


Figure~\ref{fig:bw} and \ref{fig:bwring} compare the training speed of baseline and ByteScheduler with \name on three models under different network bandwidth ranging from 500Mbps to 25 Gbps (500Mbps, 1Gbps, 5Gbps, 10Gbps, 25Gbps) under two different architectures (PS $\&$ All-reduce), which are implemented by MXNet and PyTorch respectively.

 From Figure~\ref{fig:bw} (a)-(c) We observe that \name can outperform baseline / ByteScheduler with BO by 16.7$\%$ to  60.1$\%$ / 3.1$\%$ to 17.5$\%$ for ResNet50, 30.2$\%$ to  94.4 $\%$ / 15.5$\%$ to 33.2$\%$ for VGG16, 40.7$\%$ to 79.7$\%$ / 10.6$\%$ to 31.1$\%$ for AlexNet. 
\name's meta network takes into consideration of the model's layer-wise information and dynamic change of available resources. This is the reason \name outperform baseline and ByteScheduler (BO) .
 We could observe the same trend for Ring all-reduce in PyTorch in Figure~\ref{fig:bwring} (a)-(c), 15.1$\%$ to 144.4$\%$ improvements for baseline, and 2.3$\%$ to 17.4 $\%$ improvements for ByteScheduler. This shows that \name is not only adapt to different run-time environments, but also generic to different frameworks.


\subsection{Reconfiguration Overhead and Speed}\label{subsec:deep}
\begin{figure}
	\centering
	\includegraphics[width=0.45\textwidth]{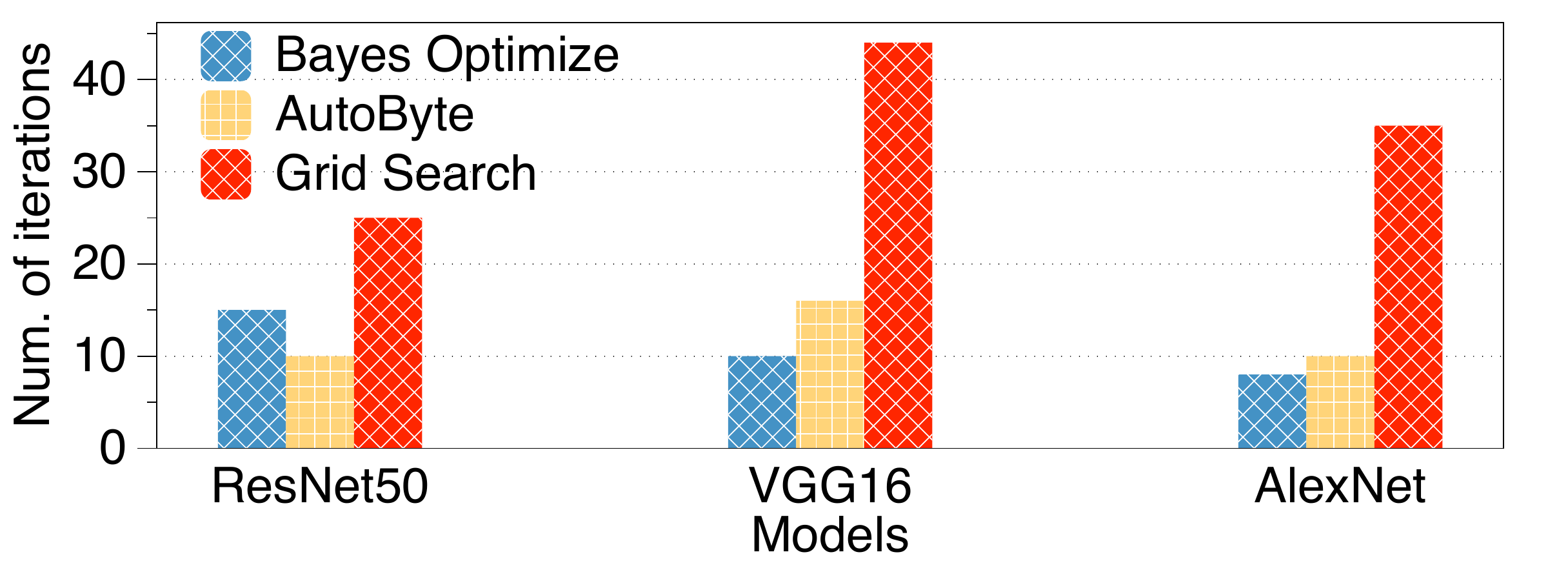}
	\caption{Search costs of different searching algorithms}
	\label{fig:cost}
\end{figure}

To make \name ready deployable, we need to ensure a low overhead and fast execution speed. 
To measure the overhead, we compare \name with a constant hyper-parameter setting. We find that their cpu utilization is almost the same (difference is less than 1\%). 
To measure the execution speed, we compare \name with grid search and BO, as shown in Figure~\ref{fig:cost}, \name reduce the execution time by $28\% \sim 81\%$ compared to the grid search and the performance is close to BO.

    \section{Related work}\label{sec:related}

\par \noindent \textbf{Communication optimization for distributed training.} Generally speaking, there are a wide range of approaches we can explore to optimize the communication for distributed DNN training. These include, but are not limited to: 1) using large mini-batch~\cite{goyal2017accurate} and periodic communication~\cite{localsgd} to reduce the communication rounds; 2) using gradient compression technique, e.g., gradient sparsification~\cite{lin2017deep} and quantization~\cite{alistarh2017qsgd}, to reduce the taffic volume in each iteration; 3) relaxing the synchronization requirement~\cite{ssp,gaia,wang2020divide}; 4) taking the intra-machine GPU topology into consideration~\cite{byteps}; 5) designing a parameter exchanging scheme considering the network topology~\cite{wan2020rat}; 6) overlapping communication with computation~\cite{poseidon,bytescheduler,p3}; 7) leveraging advanced communication library, e.g., ZMQ~\cite{zmq} and NCCL~\cite{nccl}; 8) exploiting fast network protocols, e.g., RDMA~\cite{rdma,yi2017towards}; 9) performing in-network aggregation to reduce the in-network traffic volume~\cite{switchml,atp,chen2018programmable}; 10) minimizing network flow completion time by using congestion control~\cite{alizadeh2010data,chen2013towards}, flow scheduling~\cite{bai2015information,li2017rate} or coflow scheduling~\cite{chowdhury2014efficient,zhang2016coda,susanto2016stream,zhao2015rapier}. We note that, while some of these methods have already been integrated into distributed DNN training systems, others remain to be explored in the future.

\vspace{1mm}
\par \noindent \textbf{Automatic parameter configuration.} Parameter configuration is necessary for many applications, such as big data analytics~\cite{alipourfard2017cherrypick}, tuning machine learning hyper-parameters~\cite{snoek2012practical}, and databases~\cite{duan2009tuning}. Recently, many works focus on automatic configure the parameters. For datacenter congestion control, PCC~\cite{dong2015pcc} leverages the reinforcement learning to automatically set the congestion windows. For the flow scheduling, AuTO~\cite{chen2018auto} dynamically chooses the suitable priority for each flow. For hyper-parameters tuning of machine learning, AutoML~\cite{he2021automl} can automatically find the optimal setting.

    \section{Conclusion}
In this paper, we have presented a methodology to automatically tune parameter configuration of ByteScheduler communication scheduling system. We designed \name, which is composed of Meta Network Optimizer, Optimization Trigger and Elastic Execution, to dynamically select the two critical parameters, the partition size and the credit size. The Meta Network Optimizer estimates the optimal configuration using runtime metrics. Following the decision of Optimization Trigger, the Elastic Execution applies the necessary changes dynamically to partition size and credit size without stopping the running job. Thus, \name can further increase the scheduling efficiency and decrease the training time. Our evaluation shows that \name frees ML developers of choosing the right parameter configuration by tuning the system configuration automatically.\name can optimize the system configuration when resource availability changes, reducing 33.2\% training time in dynamic network environment. 

\bibliographystyle{plain}
\bibliography{ref}
\end{document}